\documentclass[conference]{IEEEtran}
\usepackage{newtxtext}
\usepackage[T1]{fontenc}
\usepackage[utf8]{inputenc}
\usepackage[english]{babel}
\usepackage{microtype}
\usepackage{graphicx}
\usepackage{booktabs}
\usepackage{csquotes}
\usepackage{float}
\usepackage{nicefrac}
\usepackage{xspace}
\usepackage{enumitem}
\usepackage{wrapfig}
\usepackage{multirow}
\usepackage[svgnames]{xcolor} 
\usepackage[colorlinks=true, linkcolor=Navy, citecolor=Navy]{hyperref}
\usepackage{tabularray}

\usepackage{dirtytalk}

\usepackage{url}

\usepackage{amsmath}
\usepackage{amssymb}
\usepackage{mathtools}
\usepackage{amsthm}
\usepackage{algorithm}
\usepackage{algorithmic}
\usepackage[inkscapelatex=false]{svg}
\usepackage[capitalize,noabbrev,nameinlink]{cleveref}

\usepackage[sort,numbers]{natbib}

\usepackage{svg}
\usepackage{tabularx} 
\usepackage{adjustbox}
\usepackage{booktabs}
\usepackage{makecell}
\usepackage{amsthm}
\newtheorem{definition}{Definition}

\usepackage{amsmath,amsfonts,bm}

\def\eqref#1{equation~\ref{#1}}

\def\1{\bm{1}}

\DeclareMathAlphabet{\mathsfit}{\encodingdefault}{\sfdefault}{m}{sl}
\SetMathAlphabet{\mathsfit}{bold}{\encodingdefault}{\sfdefault}{bx}{n}

\usepackage{url}

\newcommand{\epsdel}{\ensuremath{(\varepsilon,\delta)}\xspace}
\newcommand{\epsdeleps}{\ensuremath{(\varepsilon,\delta(\varepsilon))}\xspace}
\newcommand{\epsdelepsi}{\ensuremath{(\varepsilon_i,\delta_i(\varepsilon_i))}\xspace}
\newcommand{\epsdeli}{\ensuremath{(\varepsilon_i,\delta_i)}\xspace}
\newcommand{\epsdeldel}{\ensuremath{(\varepsilon_i,\delta_i,\overline{\Delta})}\xspace}

\title{Your Privacy Depends on Others: \\ Collusion Vulnerabilities in Individual Differential Privacy \thanks{This work has been accepted for publication at the IEEE Conference on Secure and Trustworthy Machine Learning (SaTML). The final version will be available on IEEE Xplore.}}

\IEEEoverridecommandlockouts
\author{
\IEEEauthorblockN{
        Johannes Kaiser\textsuperscript{1, 2}, 
        Alexander Ziller\textsuperscript{1}, 
        Eleni Triantafillou\textsuperscript{4}, 
        Daniel Rückert\textsuperscript{1, 3}, 
        and Georgios Kaissis\textsuperscript{2, *}\thanks{\textsuperscript{*}Work done while at Google DeepMind.}
    }
    \IEEEauthorblockA{
        \textsuperscript{1}Chair for AI in Healthcare and Medicine, Technical University of Munich and TUM University Hospital, Munich, Germany\\
        \textsuperscript{2}Chair for Human-centred Transformative AI, HPI, University of Potsdam, Germany\\
        \textsuperscript{3} Department of Computing, Imperial College London, UK \\
        \textsuperscript{4}Google DeepMind\\
         Email: johannes.kaiser@tum.de
    }
}

\begin{document}

\maketitle

\begin{abstract}
Individual Differential Privacy (iDP) promises users control over their privacy, but this promise can be broken in practice. 
We reveal a previously overlooked vulnerability in sampling-based iDP mechanisms: while conforming to the iDP guarantees, an individual's privacy risk is not solely governed by their own privacy budget, but critically depends on the privacy choices of all other data contributors.
This creates a mismatch between the promise of individual privacy control and the reality of a system where risk is collectively determined.
We demonstrate empirically that certain distributions of privacy preferences can unintentionally inflate the privacy risk of individuals, even when their formal guarantees are met.
Moreover, this excess risk provides an exploitable attack vector.
A central adversary or a set of colluding adversaries can deliberately choose privacy budgets to amplify vulnerabilities of targeted individuals. 
Most importantly, this attack operates entirely within the guarantees of DP, hiding this excess vulnerability.
Our empirical evaluation demonstrates successful attacks against 62\% of targeted individuals, substantially increasing their membership inference susceptibility.
To mitigate this, we propose \epsdeldel-iDP a \textit{privacy contract} that uses $\Delta$-divergences to provide users with a hard upper bound on their \textit{excess vulnerability}, while offering flexibility to mechanism design.
Our findings expose a fundamental challenge to the current paradigm, demanding a re-evaluation of how iDP systems are designed, audited, communicated, and deployed to make excess risks transparent and controllable. 
\end{abstract}


\section{Introduction}
The right to privacy is fundamentally personal with individuals known to exhibit varying privacy preferences \citep{Jensen2005Privacy, Berendt2005Privacy, taylor2003most, chen2025role}. 
Crucially, privacy is understood as a context-dependent right humans exercise \textit{individually}, that is, independent of the choices made by others \citep{kan2023seeking}.
A central challenge of privacy-preserving machine learning has been translating this personal right into a technically robust and verifiable guarantee. 
While Differential Privacy (DP) \citep{dwork2006calibrating} has emerged as the gold standard for providing robust privacy guarantees, typically a single DP budget (as parametrized by an \epsdel-tuple) applies to all individuals in the database. 
Failing to account for the diverse privacy preferences of individuals, this falls short of the ideal.

To address this, the paradigm of individual DP (iDP) has been created.
iDP promises to empower data contributors by allowing them to align their personal privacy preferences with formal guarantees.
Several iDP strategies have been proposed, including accounting mechanisms, such as odometers and filtering, and assignment mechanisms. 
While accounting mechanisms track the consumed privacy budget and remove samples once their budget is exhausted, assignment mechanisms set training parameters on a per-sample basis, ensuring that each individual budget is depleted only when the algorithm concludes.
While filtering \citep{feldman2022individual, koskela2022individual} naturally extend standard non-individual DP formulations of privacy, it suffers from catastrophic forgetting \citep{koskela2022individual}, potentially diminishing the utility of the mechanism.
Odometers track, but do not enforce sample-level privacy preferences \citep{yu2022individual, whitehouse2023fully, lécuyer2021practical, rogers2016privacy}. 
Consequently, while valuable for transparently reporting privacy costs, they are a mismatched tool for upholding individual privacy preferences.

\begin{figure}[htbp]
    \centering
    \includesvg[width=1\linewidth]{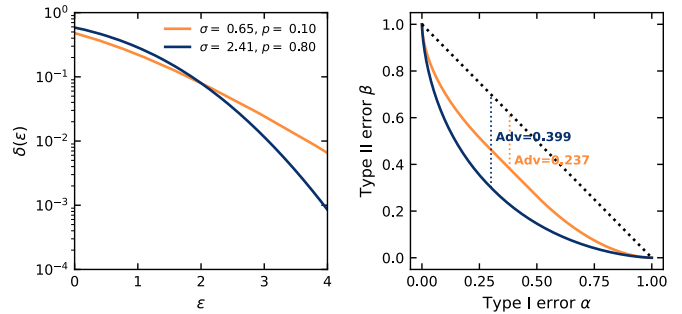}
    \caption{
    Privacy profiles (left) and adversarial-advantage trade-off curves (right) for subsampled Gaussian mechanisms with varying sampling rates and noise multipliers.
    While all are calibrated to $(2, 0.08)$-DP, their privacy profiles and thus the adversarial advantage differ substantially, leading to different levels of protection and distinct vulnerabilities. 
    }
    \label{fig:excessvul}
\end{figure}
Thus, assignment-based methods have shown superior utility and practical applicability in providing individual privacy guarantees \citep{boenisch2023have, koskela2022individual}. 
Assignment-based methods \citep{boenisch2023have, jorgensen2015conservative, heo2023personalized} determine sample-level parameters of an algorithm, such that the resulting method yields a desired iDP guarantee for each sample.
Two domains of assignment-based iDP-based methods have emerged, either adapting the sampling rates of individual data points, leveraging the subsampling amplification of DP \citep{kasiviswanathan2011can, abadi2016deep, wang2019subsampled}, or sensitivity-based leveraging sample-level sensitivity bounds to yield iDP-compliant mechanisms.
Within these two implementations of assignment-based iDP, sampling-based iDP has empirically with higher utility \citep{boenisch2023have}.
In sampling-based iDP, the sampling rates of each data point are calibrated to fulfill the individual privacy preferences.
Exemplarily, if a group of users chooses a stricter privacy budget, their sampling rates decrease to maintain their desired individual privacy guarantees. 
However, this paradoxically leads to sample interdependence: to maintain a fixed batch size for training, the sampling rate of other data points in the dataset has to change accordingly.
Therefore, what began as a promise of individual control paradoxically creates a system where \textit{one person's privacy protecting is influenced by the choices of others}. 
Explicitly, while individuals demand individual privacy protection through individual budgets, their privacy risk is collectively determined.

The specific vulnerability in sampling-based iDP methods stems from calibrating the budget to a single \epsdel. 
This single-point calibration is an incomplete specification of an algorithm's true vulnerability, as it solely describes the behaviour of the mechanisms at this calibration point.
Thus, despite identical nominal guarantees, differently calibrated mechanism parameters (i.e.\@, the sampling rate) can expose data contributors to vastly differing real-world risks as previously shown, e.g.\@, in \citet{kaissis2024beyond, lokna2023group, hayes2023bounding} and evidenced in \autoref{fig:excessvul}.
In brief, in sampling-based iDP assignment, changes in the sampling rate to accommodate individual privacy preferences ultimately result in the use of different privacy-preserving mechanisms. 
While providing the desired \epsdel privacy guarantees at the calibration point, this leads to an individual's actual privacy protection becoming \textit{silently dependent} on the choices of all other data contributors.

We analyze this excess vulnerability from two perspectives. 
First, we provide \textit{a theoretical analysis} that the interdependence creates a channel for privacy degradation. 
Second, we present \textit{empirical evidence} demonstrating that this effect manifests itself in practical machine learning scenarios, leading to a significant increase in the vulnerability for selected data points. 

We show that this interdependence of the distribution of privacy preferences and individual protection creates an exploitable attack vector for adversarial attacks. 
By coordinating their own privacy budgets, a group of colluding adversaries or a central entity can strategically manipulate the mechanism's parameters to increase the excess risk of a targeted individual, placing the targeted individual at a higher privacy risk.

The main contributions of our work are as follows:
\begin{itemize}
    \item We are the first to identify and formalize that sampling-based iDP mechanisms violate their core promise of individual privacy control by creating systematic interdependencies between users' privacy risks. 
    Specifically, we show that individual sampling rates are coupled through shared mechanism parameters, making each user's excess vulnerability a function of the entire dataset's privacy budget distribution rather than solely their personal choice.
    \item We introduce two novel attacks exploiting the privacy interdependence while operating within DP's formal guarantees: (1) \textit{Budget Manipulation Attack}, where a central entity strategically assigns privacy budgets to maximize target vulnerability, and (2) \textit{Collusion Attack}, where coordinating participants collectively amplify a victim's risk without centralized control. 
    Our evaluation shows successful attacks against 62\% of targeted individuals.
    \item Finally, we propose using $\Delta$-divergences \citep{kaissis2024beyond} to upper-bound the resulting excess vulnerabilities and the use of \epsdeldel-DP. This new privacy guarantee extends DP to provide a formal, transparent bound on the excess vulnerability. This enables transparent communication of actual privacy risks while preserving flexibility for practical mechanism design, addressing the fundamental gap between formal guarantees and individual protection.
\end{itemize}

\section{Background and Related Work}
\subsection{Differential Privacy}
Differential Privacy (DP) adopts an adversarial view of an algorithm's output distribution. 
Intuitively, if the inclusion or exclusion of a single data point does not significantly alter the output distribution, an adversary cannot confidently infer the presence of an element in the dataset. 
Formally, DP quantifies this notion of indistinguishability using a pair of parameters \epsdel (termed the privacy budget), which bounds how much the output distribution can change when a single data point is added, removed, or replaced ( $D^{-i},  D^{+i}, D^{\sim i}$ respectively).
\begin{definition}[Approximate DP, \epsdel-DP, \citep{dwork2006calibrating}]
\label{th:approxDP}
    A randomized algorithm $M: Z^* \rightarrow O$ satisfies \epsdel-DP for a fixed $\varepsilon$ and $\delta$ under removal, addition, or replacement, if for all databases $D \in Z^*$ and its neighbouring databases $D'$ where $D'$ is one of $D^{-i},  D^{+i}, D^{\sim i}$ respectively for all data points $z_i$ and for all measurable sets $S \subseteq O$,
    \[\mathrm{Pr}[M(D) \in S] \leq e^{\varepsilon} \mathrm{Pr}[M(D') \in S] + \delta\]
\end{definition}


The Gaussian mechanism provides a principled way to achieve DP by adding calibrated noise to function outputs. For a function $f: Z^* \to \mathbb{R}^d$, the mechanism operates by perturbing the actual output with Gaussian noise scaled according to the function's sensitivity and desired privacy parameters.

\begin{definition}[Gaussian Mechanism for \epsdel-DP]
\label{def:gaussian_mechanism}
Let $f : Z^* \to \mathbb{R}^d$ be a function with global $\ell_2$-sensitivity $\Delta_f = \max_{D, D'} \|f(D) - f(D')\|_2$ over all pairs of neighbouring databases $D$ and $D'$ where $D'$ is one of $D^{-i}, D^{+i}, D^{\sim i}$ for all data points $z_i$. 
For any $\varepsilon \geq 0$ and $\delta \in [0,1]$, the Gaussian mechanism
\[
M(D) = f(D) + \mathcal{N}(0, \sigma^2 I)
\]
satisfies \epsdel-DP if and only if
\[
\Phi\left(\frac{\Delta_f}{2\sigma} - \frac{\varepsilon \sigma}{\Delta_f}\right) - e^\varepsilon \Phi\left(-\frac{\Delta_f}{2\sigma} - \frac{\varepsilon \sigma}{\Delta_f}\right) \leq \delta,
\]
where $\Phi$ denotes the cumulative distribution function of the standard normal distribution.
\end{definition}

A prominent application of the Gaussian mechanism is in differentially private stochastic gradient descent (DP-SGD), a method for training machine learning models with DP guarantees.
DP-SGD has become the standard algorithm for training deep learning models with formal privacy guarantees, balancing utility and privacy in high-dimensional optimization problems \citep{Song2013StochasticGD, abadi2016deep}.
DP-SGD is operationalized by adding calibrated noise to the contributions (i.e., gradients) of individual samples to bound the signal of these samples and consequently their contribution.
To add sufficient noise to fulfil an \epsdel-DP guarantee, the magnitude of the noise has to be calibrated to the maximal sensitivity of the algorithm.
As for SGD, the gradients are not bounded, thus a gradient norm bound is imposed with a predefined magnitude to limit the sensitivity.
Furthermore, the privacy guarantees of DP-SGD can be enhanced through the subsampling amplification property of DP, which applies when an algorithm operates on batches with each batch generated by independently including data points via Poisson sampling \citep{kasiviswanathan2011can, abadi2016deep}. 
Since the optimizer only sees a random subset of the data for each gradient update, the per-example privacy loss is significantly reduced. 
A key limitation of classical DP-SGD is its uniform treatment of privacy requirements across all individuals in the dataset. 
This one-size-fits-all approach may be overly restrictive for some participants while potentially insufficient for others who require stronger privacy protection. 
This motivates the need for more flexible privacy frameworks that accommodate heterogeneous privacy preferences.

\subsection{Individual Differential Privacy}
Individual Differential Privacy (iDP) extends the classical notion of DP by allowing individualized privacy guarantees for each data point.
Rather than applying a uniform privacy budget to all individuals in the dataset, iDP accounts for the heterogeneity of privacy preferences by assigning personalized bounds on information leakage.
Similarly to conventional DP, iDP can be expressed in terms of Rényi divergences \citep{feldman2022individual} and hypothesis testing frameworks \citep{koskela2022individual}; we will adopt the following definition:

\begin{definition}[Approximate Individual DP, \citep{heo2023personalized}]
\label{def:set2}
    In the context of a privacy specification $\Psi$ and a universe of users/ data-instances $\mathcal{U}$, a randomized mechanism $\mathcal{M}:\mathcal{Z^*} \rightarrow \mathcal{O}$ satisfies $(\Psi, \Omega)$-individual DP (($\Psi$, $\Omega$)-iDP), if for all databases $D \in Z^*$ and its neighboring databases $D' \in Z^*|D' \in \{D^{-i}  D^{+i}, D^{\sim i}\}$, and for all measurable sets $S \subseteq O$,
    \[\mathrm{Pr}[\mathcal{M}(D) \in S]\leq e^{\varepsilon_i} [\mathcal{M}(D') \in S]+ \delta_i\]
    where $\varepsilon_i\in\Psi$ and $\delta_i\in\Omega$ are the personalized privacy parameters chosen \emph{a priori} by user/for data point~$x_i$.
\end{definition}

 Individual Differential Privacy (iDP) mechanisms fall into two broad families: privacy-accounting and privacy-assignment methods. 
 Accounting-based iDP tracks each record's cumulative privacy expenditure by privacy trackers and odometers \citep{yu2022individual, whitehouse2023fully, lécuyer2021practical, rogers2016privacy}, and filters, i.e., removing any record that would exceed its predefined budget before a subsequent operation would conclude \citep{rogers2016privacy, feldman2022individual, koskela2022individual}.
 While pure budget tracking gives fine-grained visibility into each record's privacy consumption, it offers no means for data providers to set or adjust their budgets to comply with their privacy preferences. 
 Privacy filtering, by contrast, enforces individual limits by dropping data points that may exceed their privacy budget with the following computation. 
 However, this comes at a steep cost: as records get filtered out over time, models suffer from catastrophic forgetting and overall utility plummets \citep{koskela2022individual}. 
 Consequently, filtering-based iDP remains largely impractical for real-world applications.

Privacy‑assignment methods allow each user to declare their own desired privacy budgets in advance and can be integrated into learning algorithms without incurring significant utility loss when compared to non-individual DP. 
Initially proposed by \citet{jorgensen2015conservative}, these methods have been operationalized in machine learning either by leveraging the subsampling theorem of DP‑SGD \citep{kasiviswanathan2011can, heo2023personalized, boenisch2023have} or by modulating per‑sample sensitivity via sample-level individualized clipping bounds \citep{boenisch2023have}. 
This work focuses on the sampling-based approach proposed in \citep{boenisch2023have}, as \citep{heo2023personalized} mainly employs subsampling in terms of an iterative filtering strategy, which does not fully expend the privacy budgets.
We introduce sensitivity-based iDP in \Cref{app:sens_based}.

\begin{definition}[Sampling with pre-defined expected batch size, \citep{boenisch2023have}]
\label{def:individual_sampling}
    Given per-group target RDP (Rényi DP \citep{Mironov_2017}) budgets $\boldsymbol{\rho} = \{\rho_1, \ldots, \rho_M\}$ for groups $\{g_1, \ldots, g_M\}$, target $\delta$, iterations $I$, total data points $N$, group sizes $\{|\mathcal{G}_1|, \ldots, |\mathcal{G}_M|\}$, and optimal non-individual DP sampling rate $p$, the combination of noise parameter $\sigma$ and group specific sampling rates $\{p_1, \ldots, p_M\}$ are chosen such that:
    \begin{align}
        p &\approx \sum_{i=1}^M \frac{|\mathcal{G}_i|}{N} p_i \label{eq:avg_sampling_rate}\\
        \rho_i &\leq I \cdot 2p_i^2 \frac{\alpha}{\sigma^2} \quad \text{for all } i \in \{1, \ldots, M\} \label{eq:rdp_constraint}
    \end{align}

    Then the mechanism satisfies $(\alpha, \rho_i)$-RDP for group $g_i$. 
    Using the optimal RDP to \epsdel-DP conversion \citep{zhu2022optimalaccountingdifferentialprivacy}, this yields $(\varepsilon_i, \delta)$-iDP guarantees for each group $g_i$.
\end{definition}
We remark that this sampling-based individual privacy assignment was originally formulated based on groups of data sharing the same privacy budget; it can be extended to the individual level by setting the group size to one.
The key constraint we will investigate is Equation (1), which fixes the expected batch size for training stability. 
To satisfy this, if a large group of users chooses a low privacy budget $\rho_i$ (requiring a low sampling rate $p_i$), the mechanism must compensate by increasing the sampling rates of other users. 
This directly links every user's sampling probability—and thus their privacy—to the choices of others.
The individual sampling rates and the noise multiplier are determined by binary search. We refer the reader to \citep{boenisch2023have} for concrete implementation details.
\textit{Sampling with pre-defined expected batch size} defines data point-level sampling rates and a clipping norm. 
While the sampling rates are calibrated such that the mechanism satisfies $(\Psi, \Omega)$-iDP, their values largely depend on the distribution of privacy budgets across the dataset.
We discuss how to relax the constraint of a predefined expected batch size and its implications in \Cref{app:fixed_batch_size}.

\subsection{Excess Vulnerability of DP}
In the following, we will refer to \textit{excess vulnerability} and \textit{(excess) risk} as adversary advantage (i.e. membership-inference advantage) not captured by \epsdel-DP, which, by unified f-DP bouds, also upper-bounds excess risk for re-identification, attribute inference, and reconstruction \cite{kulynych2025unifying}.
Commonly, while DP mechanisms are calibrated to satisfy a specific \epsdel-guarantee, most relevant DP mechanisms (including but not limited to the Gaussian and Laplace mechanisms) satisfy DP across a continuum of \epsdeleps-values.
This continuum can be expressed via \textit{privacy profiles} \citep{balle2020privacy} or \textit{trade-off functions} \citep{dong2022gaussian}.
A single \epsdel tuple describes merely a single point on the privacy profiles and thus provides only a lossy and often misleading single-point representation of the actual privacy protection, leaving the rest of the curve unconstrained.
Critically, the privacy protection offered by a privacy-preserving mechanism depends on the complete privacy profile rather than the calibration point \citep{kaissis2024beyond}, and mechanisms with the same \epsdel guarantee may differ substantially in the protection they provide. 
This gives rise to excess vulnerability of the differentially private mechanisms not captured by the \epsdel formulation.
This phenomenon is exemplified in \autoref{fig:excessvul}: while both shown mechanisms satisfy $(2,0.08)$-DP, they exhibit distinct privacy profiles and risk (as demonstrated by the different adversarial advantages of the two mechanisms), exposing excess vulnerabilities that uniform \epsdel-DP guarantees fail to capture.

\subsection{Auditing Data Vulnerability in Machine Learning}
To assess the vulnerability of individual data samples, we conduct a privacy audit of trained models.
Specifically, we want to estimate the sample-level advantage an adversary can gain - that is, how likely an adversary can infer sensitive information about the data by observing the output of the privacy-preserving mechanism. 
Auditing techniques aim to estimate the privacy risk associated with specific data points, either by leveraging naturally occurring data or by crafting adversarial inputs \citep{nasr2021adversary, steinke2023privacy, mahloujifar2024auditing, boglioni2025optimizing}.
While the latter is beneficial for tightly auditing the privacy guarantees of an algorithm, we will focus on the former as we are interested in auditing the vulnerability of the underlying training data in a realistic scenario \citep{feldman2020neural}.
Moreover, we adopt a realistic black-box setting,  where the auditor evaluates a deployed neural network without access to its internal gradients, parameters, or training dynamics \citep{shokri2017membership, salem2018ml, yeom2018privacy, sablayrolles2019white, song2021systematic, hui2021practical, boglioni2025optimizing, carlini2022membership, zarifzadeh2024low}.
This assumption aligns with many practical scenarios where deployed machine learning models are exposed via APIs, but their internal workings remain proprietary or inaccessible \citep{hisamoto2020membership, truex2019demystifying}.

Within this black-box framework, one of the most widely studied and practical tools for auditing is the membership inference attack (MIA).
In the context of membership inference attacks (MIAs), the objective of the auditor is to determine whether a given data point $x$ was part of the training dataset $D_{\text{train}}$ of a machine learning model \citep{carlini2022membership, zarifzadeh2024low}. This problem can be formalized as a binary hypothesis testing task, where the two competing hypotheses are defined as:
\begin{align*}
    H_0 &: x \notin D_{\text{train}}, \\
    H_1 &: x \in D_{\text{train}}.
\end{align*}

Under the null hypothesis $H_0$, the data point $x$ is assumed to be drawn from the general population or from a hold-out test set, and therefore has not been observed by the model during training. 
In this case, the model's output on $x$ reflects its ability to generalize rather than memorize the training data. 
In contrast, under the alternative hypothesis $H_1$, the point $x$ belongs to the training dataset, and the model may exhibit behaviours indicative of having interpolated the point, such as unusually confident predictions, lower loss values, or characteristic activation patterns in intermediate layers. 
The latter provides evidence to perform the hypothesis test.

Formally, a membership inference attack aims to design a decision function $\mathcal{A}$ that maps the model's outputs and potentially additional information about $x$ to a binary prediction:
\[
\mathcal{A}(x) = 
\begin{cases}
1 & \text{if } x \in D_{\text{train}} \text{ (predict } H_1\text{)}, \\
0 & \text{if } x \notin D_{\text{train}} \text{ (predict } H_0\text{)}.
\end{cases}
\]
Membership inference naturally aligns with the guarantees provided by DP. 
It is considered a strong attack, as it aims to infer just a single bit of information (member/ non-member) about the data \citep{yeom2018privacy, shokri2017membership, ye2022enhanced, carlini2022membership, sablayrolles2019white, watson2021importance, zarifzadeh2024low}.
For instance, a model may exhibit higher confidence in its predictions for member data points.

Likelihood Ratio Attack (LiRA) \citep{carlini2022membership}, an instantiation of MIA, bases the decision rule for the hypothesis test on the difference in distributions of logits (which are assumed to be Gaussian).
LiRA involves training numerous shadow models on random data subsets. 
For each target data point, it generates two distinct populations of models: IN models, trained on subsets of the initial dataset containing a specific sample, and  OUT models, trained on subsets excluding it.
This process yields two empirical distributions for the model's output logits on the target point: one conditioned on its inclusion in the training set, $P_{\text{in}}$, and another conditioned on its exclusion, $P_{\text{out}}$.

The performance of such an attack is typically evaluated in terms of the probabilities of false positives and false negatives. A false positive occurs when $\mathcal{A}(x) = 1$ despite $x \notin D_{\text{train}}$, whereas a false negative occurs when $\mathcal{A}(x) = 0$ despite $x \in D_{\text{train}}$. 
This enables the rigorous analysis of the vulnerability of machine learning models to MIA and the quantification of the privacy leakage associated with model outputs.

A standard metric to capture this vulnerability is the \textbf{MIA Advantage}, defined as the difference between the maximal True Positive Rate (TPR; $1-\beta$) and the False Positive Rate (FPR $\alpha$):
$$
\text{Adv}(\mathcal{A}) = \max_{\alpha}(1-\beta(\alpha) - \alpha)
$$
The MIA advantage captures the maximal gap between true and false positive rates, effectively summarising the adversary's best possible discriminatory power independent of decision bias.
While we will report the advantage as theoretical support for our claims, it is difficult to determine empirically on a per-sample level, as it constitutes an aggregate statistic.
Consequently, we follow \citep{carlini2022membership} and use the \textit{privacy score $\mathrm{priv}$}, defined as:
\[
    \mathrm{priv} = \frac{|\mu_{\text{in}} - \mu_{\text{out}}|}{\sigma_{\text{in}} + \sigma_{\text{out}}}
\]
with $\mu_{\text{in}}$, $ \mu_{\text{out}}$ as the means and $\sigma_{\text{in}}$, $\sigma_{\text{out}}$ as the variances if $P_{\text{in}}$ and $P_{\text{out}}$ respectively.
$\mathrm{priv}$ measures the distance between the IN and OUT distributions and serves as a good proxy for the MIA advantage.


\section{Individual Privacy Assignment leads to Excess Vulnerabilities}
Sampling-based iDP determines sampling rates and noise multipliers such that DP-SGD with these parameters satisfies each sample's target privacy budget \epsdeli, for $i$ being specific privacy groups.
As \textit{privacy groups}, we describe sets of data points sharing the same privacy budget.
Throughout the evaluation, we will report privacy groups and the proportion of the dataset they make up.
The algorithm is further constrained by a fixed expected batch size, which enforces an average sampling rate across the dataset. 

As a result, an individual's sampling probability is not independent, but rather depends on the distribution of sampling rates induced by the privacy preferences of all other samples.
For example, if a data contributor demands a small privacy budget, the sampling probability of their data will also be small.
Consequently, other samples in the dataset are assigned larger sampling rates to satisfy the expected batch size.

\begin{figure}[htbp]
    \centering
    \includesvg[width=\linewidth]{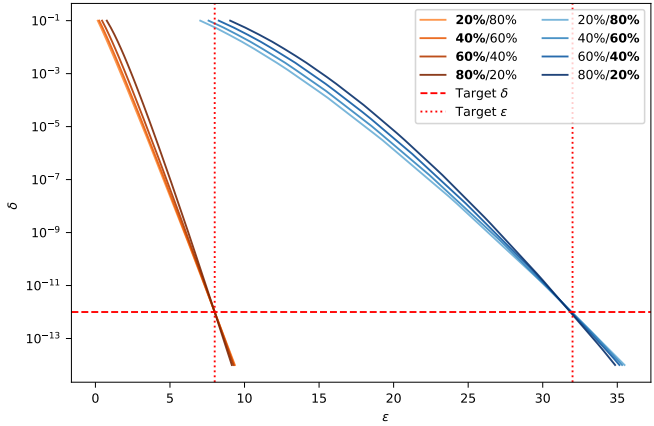} 
    \includesvg[width=\linewidth]{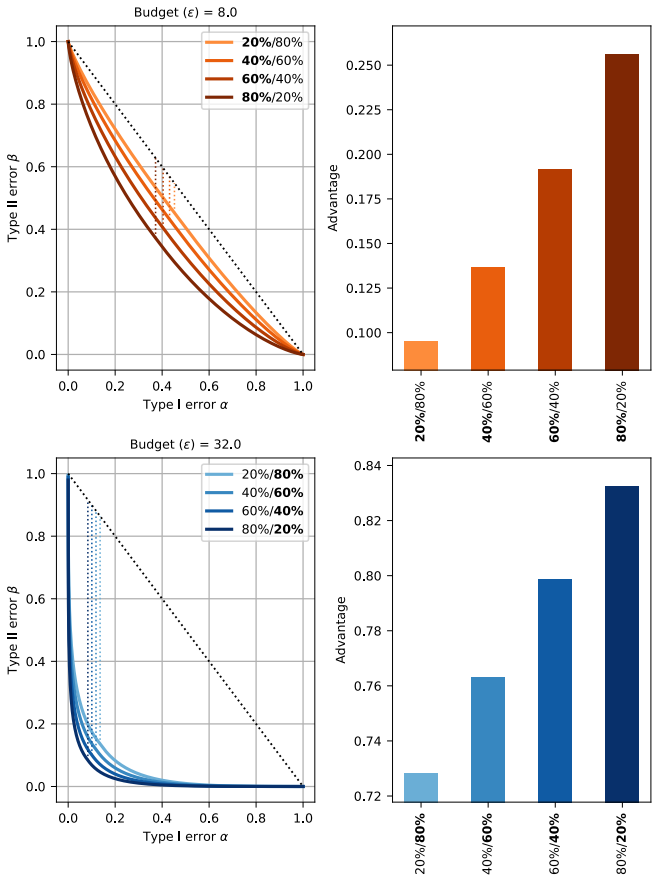} 
    \caption{Privacy profiles, trade-off functions, and adversarial advantage for a machine learning model trained with sampling-based iDP.
    While all mechanisms are calibrated to guarantee the predefined privacy budgets $\varepsilon_1 = 8, \varepsilon_2 = 32$ at $\delta=10^{-12}$, their descriptive profiles and consequently the adversarial advantage differ substantially. We obtain the advantage analytically by using the privacy profiles and trade-off functions of the subsampled Gaussian mechanism, parametrized by the sampling rate and noise multiplier. This value serves as a theoretical upper bound on the advantage that can be observed empirically.}
    \label{fig:privacy_tradeoff}
\end{figure}

While the mechanism remains calibrated to satisfy \epsdeli-iDP for each sample, the privacy profiles change due to the change in sampling rates.
As the whole privacy profile determines the excess risk, it too changes.
As illustrated in \autoref{fig:privacy_tradeoff}, while every mechanism respects the specified \epsdeli guarantees, their resulting privacy profiles \epsdelepsi—and thus per-sample vulnerabilities—can vary substantially.

Remark: The interdependence arises due to having a blanked noise multiplier for all data points, which requires the selection of individual sampling rates such that the final mechanism is iDP with the desired privacy budgets.
Enforcing a predefined expected batch size is a design choice of the original work to allow for sampling-based iDP training with similar parameters to standard DP-SGD, incorporating information from all $\varepsilon_i$ to determine a fitting global noise multiplicator.
Nonetheless, different methods of integrating this information into the noise multiplier are possible and are further discussed in \Cref{app:fixed_batch_size}.
Throughout the paper we will assume the setting of a fixed expected batch size, as it is the formulation used in the original work, and follows a nice intuition.
However, the reason is more fundamental.

Most importantly, the observed vulnerability does not stem from implementation artefacts such as numerical errors in the binary search for mechanism parameters or imprecise conversions between privacy formulations. 
Instead, it arises from a more fundamental issue: the privacy mechanism is calibrated to satisfy an incomplete specification, namely, the per-sample guarantees $\varepsilon_i$ and $\delta_i$.
This presents a critical risk. 
Because all mechanisms are explicitly designed to meet these per-sample bounds, they are compliant on paper.
However, this compliance masks excess vulnerability as it remains unreported because it lies outside the scope of what $\varepsilon_i$ and $\delta_i$ capture. 
This means models may systematically leak more information than anticipated, even when adhering to the formal privacy parameters.

\begin{figure}[htbp]
    \centering
    \includesvg[width=1\linewidth]{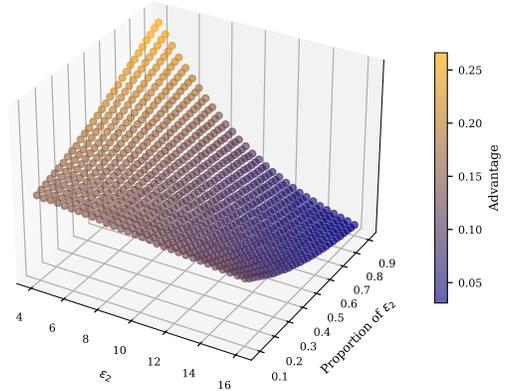}
    \caption{Adversarial advantage on a machine learning model on data points with an $\varepsilon_1 = 8, \delta=10^{-12}$ in the presence of $p$ proportion of data points of $\varepsilon_2$ when privacy is protected via sampling-based iDP. The theoretical MIA advantage and consequently the excess risk strongly depend on the proportion and privacy budget of group 2. If \textbf{$\varepsilon_2 < \varepsilon_1$}, the adversarial risk increases with decreasing $\varepsilon_2$ and increasing proportion of group 2 in the dataset. If \textbf{$\varepsilon_2 > \varepsilon_1$}, the adversarial risk decreases with decreasing $\varepsilon_2$ and also decreases the proportion of group 2 in the dataset.}
    \label{fig:3d_privacy_tradeoff}
\end{figure}

With sampling-based iDP, the extent of these excess vulnerabilities depends on the distribution of privacy parameters in the whole dataset. 
\autoref{fig:3d_privacy_tradeoff} shows the theoretical adversarial advantage for data points calibrated to $(\varepsilon_1=8,\delta=10^{-12})$, in the presence of another group of proportion size $p$ with privacy budget $\varepsilon_2$ (at $\delta=10^{-12}$). 
Given the resulting sampling rate and noise multiplier, the advantage is computed analytically from the trade-off functions of the subsampled Gaussian mechanism. 
Consequently, it constitutes a theoretical upper bound on the empirical advantage.
Vulnerability varies substantially with $\varepsilon_2$ and $p$: it increases in the presence of samples with smaller $\varepsilon_2$, and decreases with larger $\varepsilon_2$, particularly when these dominate the dataset. In this setting, the adversarial advantage ranges between $0.15$ and $0.4$.
Remark: While this may appear as a selected example, it in fact reflects a realistic machine learning training with a dataset consisting of $50{,}000$ samples, batch size $128$, over $5$ epochs - a typical training scenario for small datasets.

\section{Empirical Evaluation of the Excess Vulnerability}
\label{sec:empirical_validation}

With sampling-based iDP, excess vulnerability depends on the distribution of privacy parameters across the dataset. 
While our theoretical analysis provides an initial understanding, we will now evaluate whether these effects appear in practice. 
We thus conduct a sample-level membership inference attack to assess how the theoretically predicted vulnerabilities manifest in practice.

In this empirical study, each sample is randomly assigned a privacy budget from two predefined values, $\varepsilon_1$ and $\varepsilon_2$, such that the resulting distribution of budgets follows a prescribed distribution.
This random assignment is repeated across five seeds to ensure robustness of the results. 
We investigate the effect of this distribution by systematically varying the proportion of users with different privacy preferences.
We employ the LiRA framework to estimate per-sample logit distributions for both IN and OUT models, which are then used to quantify each data point's membership inference susceptibility.
This requires training a set of $n$ shadow models ($n = 512$ if not defined otherwise ) using sampling-based iDP.
We partition the dataset into an IN and an OUT dataset, each containing 50\% of the data while still conforming to the predefined privacy budget distribution.
Determining IN and OUT datasets with the constraint of maintaining the privacy budget distribution entails generating a biregular binary assignment matrix, which is non-trivial for random subsets.
We approximate this by weakly enforcing that each data point is present in exactly half of the datasets and strongly enforcing that the proportions remain consistent.
We further elaborate on this and the training settings in \Cref{appendix:TrainingSettings}.
The code used to generate these results is publicly available for reproducibility at \href{https://github.com/Johannes-Kaiser/Collusion-Vuln-in-DP}{https://github.com/Johannes-Kaiser/Collusion-Vuln-in-DP}
.


LiRA lacks the precision to audit per-sample risk tightly.
To isolate and measure the effect of privacy interdependence, we design an experimental setting that reduces the effect of other confounding factors (i.e., noisy measurement via LiRA).
As discussed in \Cref{app:delta}, using low $\delta$ values, we operate in a regime where the interdependence is emphasized, such that LiRA operates on a cleaner signal, allowing for a more precise measurement of the change in vulnerability attributable to the iDP mechanism itself.
Nonetheless, this deliberate choice is solely to increase the signal we intend to measure, as the excess risk in theory also appears with smaller $\delta$ values (albeit with smaller magnitude).

\subsection{Results and Observations}

Shown in \autoref{tab:empirical_priv_images}, the theoretically discussed change in vulnerability also appears in practice. 
We consider datasets with a sample-level privacy budget within two distinct privacy groups of budgets $\varepsilon_i$, at a fixed $\delta_i = 10^{-12}$.
Using the sampling-based iDP mechanism yields sampling rates and noise multiplier such that the subsequent DP-SGD mechanism conforms to the predefined iDP guarantees.
However, as discussed in the previous section, depending on the proportion these two privacy groups make up in the dataset, the sampling-based iDP mechanism yields sampling rates and noise multipliers such that the privacy profiles, as well as the trade-off functions and adversarial advantage, differ significantly.
\autoref{tab:empirical_priv_images} illustrates the distributions of empirically measured privacy scores ($\mathrm{priv}$) for the two distinct privacy groups.
With an increasing portion of data with larger privacy budgets, $\mathrm{priv}$ increases, indicating an increase in MIA vulnerability.
Consequently, even though data contributors provide their data while being guaranteed a privacy guarantee of $(\varepsilon_i, 1\times10^{-12})$, their actual privacy risk is strongly influenced by the presence and privacy guarantee of all samples in the dataset.
For further details on the training setting and datasets specifications, we refer the reader to \Cref{appendix:TrainingSettings}.


Across all datasets, the privacy score and consequently the MIA susceptibility for both privacy groups increase when there is a smaller group with a higher privacy budget. These results underscore the critical importance of considering the entire privacy parameter distribution when assessing the security of iDP systems.

\begin{table*}[!htbp]
\centering
\caption{This table presents the \textbf{empirical privacy scores} for various datasets, models, and privacy budgets as measured through membership inference. A clear trend emerges: \textbf{the privacy score consistently increases as the proportion of Group 1 with a smaller privacy budget increases}. To formally test this trend, we used the \textbf{Jonckheere-Terpstra significance test} to check for a monotonically increasing relationship between the privacy score and the proportion of Group 1. The statistical significance levels are denoted as follows: \textbf{[***]} for $p \leq 0.001$ (very highly significant), \textbf{[**]} for $p \leq 0.01$ (highly significant), \textbf{[*]} for $p \leq 0.05$ (significant), \textbf{[]} for $p > 0.05$ (not significant). In addition to the significance levels, we also report the \textbf{effect size} to quantify the magnitude of the observed effect. Our results demonstrate a \textbf{statistically significant increase in privacy scores}, which implies a greater advantage for potential adversarial attacks.}
\label{tab:empirical_priv_images}
\adjustbox{max width=\textwidth}{%
\begin{tabularx}{\textwidth}{c c c c c X X}
\toprule
\textbf{Dataset} & \textbf{$\varepsilon_1$} & \textbf{$\varepsilon_2$} & \makecell[c]{\textbf{Significance} \\ \textbf{(JT) of group 1}} & \makecell[c]{\textbf{Significance} \\ \textbf{(JT) of group 2}} & \textbf{Privacys score for group 1} & \textbf{Privacys score for group 2} \\
\midrule
\makecell[t]{Credit Card Defaut} & 4 & 20 & \makecell[t]{[***] \\ (\textbf{ES}: 20.2)} & \makecell[t]{[***] \\ (\textbf{ES}: 17.5)} & \raisebox{-.5\height}{\includesvg[width=1\linewidth]{figures/priv_evals_box_new/boxplot_single_priv_credit_card_default_0_no_labels.svg}} & \raisebox{-.5\height}{\includesvg[width=1\linewidth]{figures/priv_evals_box_new/boxplot_single_priv_credit_card_default_1_no_labels.svg}} \\
\makecell[t]{German Credit} & 4 & 16 & \makecell[t]{[***] \\ (\textbf{ES}: 19.0)} & \makecell[t]{[***] \\ (\textbf{ES}: 17.29)} & \raisebox{-.5\height}{\includesvg[width=1\linewidth]{figures/priv_evals_box_new/boxplot_single_priv_german_credit_0_no_labels.svg}} & \raisebox{-.5\height}{\includesvg[width=1\linewidth]{figures/priv_evals_box_new/boxplot_single_priv_german_credit_1_no_labels.svg}} \\
\makecell[t]{MNIST} & 4 & 16 & \makecell[t]{[] \\ (\textbf{ES}: -)} & \makecell[t]{[] \\ (\textbf{ES}: -)} & \raisebox{-.5\height}{\includesvg[width=1\linewidth]{figures/priv_evals_box_new/boxplot_single_priv_mnist_0_no_labels.svg}} & \raisebox{-.5\height}{\includesvg[width=1\linewidth]{figures/priv_evals_box_new/boxplot_single_priv_mnist_1_no_labels.svg}} \\
\makecell[t]{MNIST \\{}[4, 1000]} & 16 & 50 & \makecell[t]{[***] \\ (\textbf{ES}: 21.2)} & \makecell[t]{[***] \\ (\textbf{ES}: 17.7)} & \raisebox{-.5\height}{\includesvg[width=1\linewidth]{figures/priv_evals_box_new/boxplot_single_priv_mnist_4_0_no_labels.svg}} & \raisebox{-.5\height}{\includesvg[width=1\linewidth]{figures/priv_evals_box_new/boxplot_single_priv_mnist_4_1_no_labels.svg}} \\
\makecell[t]{UCI-Isolet} & 1 & 10 & \makecell[t]{[***] \\ (\textbf{ES}: 30.6)} & \makecell[t]{[***] \\ (\textbf{ES}: 70.9)} & \raisebox{-.5\height}{\includesvg[width=1\linewidth]{figures/priv_evals_box_new/boxplot_single_priv_uci_isolet_0_no_labels_old.svg}} & \raisebox{-.5\height}{\includesvg[width=1\linewidth]{figures/priv_evals_box_new/boxplot_single_priv_uci_isolet_1_no_labels.svg}} \\
\makecell[t]{CIFAR-10} & 16 & 50 & \makecell[t]{[***] \\ (\textbf{ES}: 13.9)} & \makecell[t]{[***] \\ (\textbf{ES}: 53.5)} & \raisebox{-.5\height}{\includesvg[width=1\linewidth]{figures/priv_evals_box_new/boxplot_single_priv_cifar10_0_no_labels.svg}} & \raisebox{-.5\height}{\includesvg[width=1\linewidth]{figures/priv_evals_box_new/boxplot_single_priv_cifar10_1_no_labels.svg}} \\

\makecell[t]{OrganCMNIST \\{}[2000]} & 8 & 32 & \makecell[t]{[***] \\ (\textbf{ES}: 18.5)} & \makecell[t]{[***] \\ (\textbf{ES}: 42.2)} & \raisebox{-.5\height}{\includesvg[width=1\linewidth]{figures/priv_evals_resnet/boxplot_single_priv_organcmnist_0_no_labels_2000.svg}} & \raisebox{-.5\height}{\includesvg[width=1\linewidth]{figures/priv_evals_resnet/boxplot_single_priv_organcmnist_1_no_labels_2000.svg}} \\
\makecell[t]{OrganSMNIST \\{}[2000]} & 8 & 32 & \makecell[t]{[***] \\ (\textbf{ES}: 34.6)} & \makecell[t]{[***] \\ (\textbf{ES}: 71.8)} & \raisebox{-.5\height}{\includesvg[width=1\linewidth]{figures/priv_evals_resnet/boxplot_single_priv_organsmnist_0_no_labels_2000.svg}} & \raisebox{-.5\height}{\includesvg[width=1\linewidth]{figures/priv_evals_resnet/boxplot_single_priv_organsmnist_1_no_labels_2000.svg}} \\
\makecell[t]{Pneumonia \\{}[2000]} & 8 & 32 & \makecell[t]{[***] \\ (\textbf{ES}: 15.9)} & \makecell[t]{[***] \\ (\textbf{ES}: 14.3)} & \raisebox{-.5\height}{\includesvg[width=1\linewidth]{figures/priv_evals_resnet/boxplot_single_priv_pneumoniamnist_0_no_labels_2000.svg}} & \raisebox{-.5\height}{\includesvg[width=1\linewidth]{figures/priv_evals_resnet/boxplot_single_priv_pneumoniamnist_1_no_labels_2000.svg}} \\
\makecell[t]{HAM10k \\{}[2000]} & 8 & 32 & \makecell[t]{[***] \\ (\textbf{ES}: 31.9)} & \makecell[t]{[***] \\ (\textbf{ES}: 37.2)} & \raisebox{-.5\height}{\includesvg[width=1\linewidth]{figures/priv_evals_resnet/boxplot_single_priv_dermamnist_0_with_labels_2000.svg}} & \raisebox{-.5\height}{\includesvg[width=1\linewidth]{figures/priv_evals_resnet/boxplot_single_priv_dermamnist_1_with_labels_2000.svg}} \\
\bottomrule
\end{tabularx}
}

\end{table*}

\subsection{Are we actually measuring what we want to measure?}
To ensure that the effects we show originate from the sampling-based iDP mechanism and are not attributable to artifacts, for example, due to changing distribution in desired privacy budgets, we compare our results to sensitivity-based iDP (which we briefly introduce in \Cref{app:sens_based}) \citep{boenisch2023have} as an ablation study.
In sensitivity-based iDP, the scale of the sample-level additive noise and clipping bounds are adapted to fit the desired sample-level privacy guarantee. 
In essence, sensitivity-based iDP trades of global changes in the magnitude of the additive noise with sample-level changes in clipping bounds.
While this changes how the mechanism is operationalized, it does not change the privacy profiles and consequently the MIA risk.
This is attributed to the fact that the privacy profiles depend solely on the relation of noise to the clipping bound.
We further elaborate on this in \Cref{app:sens_based}.
\begin{table}[!htbp]
\centering
\caption{Empirical adversarial across selected datasets using sensitivity-based iDP using the same training setting as for \autoref{tab:empirical_priv_images}.
While Jonckheere-Terpstra significance test indicates statistical significance (the underlying effect is explained in \autoref{app:sens_based}) with an effect size smaller by an order of magnitude when compared to sampling-based iDP.}
\label{tab:empirical_priv_images_clipping}
\begin{tabularx}{\linewidth}{l X X}
\toprule
\textbf{Dataset} & \textbf{Privacy Scores for group 1} & \textbf{Privacy Scores for group 2}\\
\midrule
\makecell[t]{MNIST \\{}[4, 1000]} & \raisebox{-.5\height}{\includesvg[width=1\linewidth]{figures/adv_evals_clip/boxplot_single_priv_mnist_4_0_no_labels_clipping.svg}} & \raisebox{-.5\height}{\includesvg[width=1\linewidth]{figures/adv_evals_clip/boxplot_single_priv_mnist_4_1_no_labels_clipping.svg}} \\

\makecell[t]{Peripheral Blood \\{}[2000]} & \raisebox{-.5\height}{\includesvg[width=1\linewidth]{figures/adv_evals_clip/boxplot_single_priv_bloodmnist_0_no_labels_clipping.svg}} & \raisebox{-.5\height}{\includesvg[width=1\linewidth]{figures/adv_evals_clip/boxplot_single_priv_bloodmnist_1_no_labels_clipping.svg}} \\

\makecell[t]{OrganCMNIST \\{}[2000]} & \raisebox{-.5\height}{\includesvg[width=1\linewidth]{figures/adv_evals_clip/boxplot_single_priv_organcmnist_0_with_labels_clipping.svg}} & \raisebox{-.5\height}{\includesvg[width=1\linewidth]{figures/adv_evals_clip/boxplot_single_priv_organcmnist_1_with_labels_clipping.svg}} \\

\bottomrule
\end{tabularx}

\end{table}
Consequently, it allows us to compare our evaluation against a method that operates in the identical setting, providing the same \epsdeli-DP guarantee with the sole difference in how DP is enforced on a per-sample level.
\autoref{tab:empirical_priv_images_clipping} shows the empirical privacy scores using this sensitivity-based approach of iDP in the same training paradigm as used in \autoref{tab:empirical_priv_images}. 
Compared to our results on sampling-based iDP (\autoref{tab:empirical_priv_images}), even though there is a measurable trend its effect size is smaller by an order of magnitude compared to sampling based iDP.
(Most likely the trend emerges due to different model behavior with varying per-sample clipping bounds as further elaborated on in \Cref{app:sensitivit_based}).
The privacy profiles and consequently the MIA advantage, however, are independent of the privacy budget distribution for sensitivity-based iDP.
We thus conclude that the remaining change in privacy scores can be causally attributed to the interdependence in the sampling-based iDP mechanism.

\section{Exploiting iDP-based Excess Vulnerabilities in Adversarial Settings}
\subsection{Budget Manipulation Attack}
\begin{figure}[htbp]
    \centering
    \includesvg[width=0.48\textwidth]{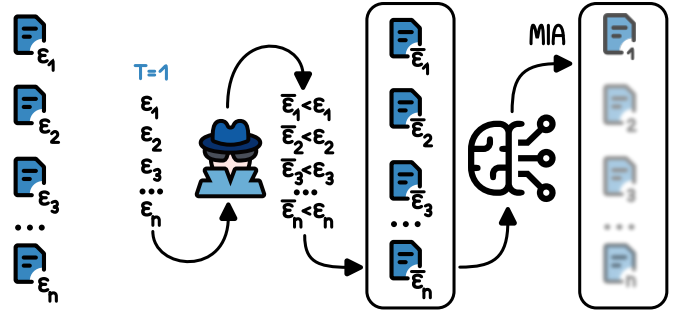} 
    \caption{Schematisation of the Budget Manipulation Attack. A central adversary (model owner) performs a computation expending at most $\varepsilon_i, \delta$ privacy budget for data point $i$. The adversary deliberately assigns sample-level privacy budgets to increase the excess risk of a targeted data point. Most importantly, the choices of privacy budget are bounded by the stipulated $\varepsilon_i, \delta$ privacy budgets.}
    \label{fig:attack1}
\end{figure}
The previously explored vulnerabilities of sampling-based individual DP assignment give rise to novel attacks, which we term the \textit{Budget Manipulation Attack}.
The threat model under which the attack operates is visualized in \autoref{fig:attack1} and defined as follows:

\begin{definition}[Threat Model: Budget Manipulation Attack]
Let $\mathcal{A}$ be a (possibly randomized) training algorithm satisfying iDP, i.e., for a dataset $D = \{x_i, \dots, x_n\}$ and corresponding privacy budgets $\{\varepsilon_i\}_{i=1}^n$, the algorithm $\mathcal{A}$ satisfies $(\varepsilon_i, \delta)$-DP for each sample $x_i$ individually.

The adversary operates under the following capabilities and constraints:
\begin{enumerate}
    \item \textbf{No Model Access.} The adversary has no access to the model $\mathcal{A}(D)$, its parameters, gradients, intermediate computations, or outputs. The attack is thus \emph{data-only} and conducted entirely before training.
    
    \item \textbf{Target Selection.} The adversary selects a target data point $x_j \in \mathcal{X}$ to increase its vulnerability to a privacy attack (e.g., membership inference).
    
    \item \textbf{Privacy Budget Control.} The adversary is permitted to assign privacy budgets $\{\varepsilon_i\}_{i=1}^n$ to all training points, subject to the constraint that for all $i \in \{1, \dots, n\}$:
    \[
    \varepsilon_i \in [0, \varepsilon_i^{\max}],
    \]
    where $\varepsilon_i^{\max}$ is a fixed upper bound on the allowable privacy budget per individual sample.
    
    \item \textbf{Attack Objective.} The adversary seeks to assign the privacy budget vector $\Psi = (\varepsilon(x_1), \dots, \varepsilon(x_n))$ to maximize the expected vulnerability of the target point $x_j$ under the training algorithm $\mathcal{A}$. Formally, the adversary solves:
    \[
    \max_{\varepsilon_i \in [0, \varepsilon_i^{\max}]^n_{i=1}} \mathbb{E}_{\mathcal{A}} \left[ V(x_j, \mathcal{A}(D)) \right],
    \]
    where $V(x_j, \mathcal{A}(D))$ denotes a quantitative measure of the vulnerability of $x_j$ (e.g., probability of correct membership inference by a downstream attacker), and the expectation is taken over the randomness of $\mathcal{A}$.
\end{enumerate}

\end{definition}

Within the described threat model, a particularly effective attacker strategy in the setting of sampling-based individual DP is to allocate privacy budgets in an imbalanced manner: assigning a high privacy budget $\varepsilon_j \approx \varepsilon_j^{\max}$ to the target data point $x_j$, while assigning very low budgets $\varepsilon_i < \varepsilon_j$ to all non-target data point $x_i$ for $i \neq j$. 
This allocation respects the iDP constraint $\varepsilon_i \leq \varepsilon_i^{\max}$ for all $i$, and ensures that the target's privacy guarantee remains formally calibrated to its maximal $(\varepsilon_j, \delta)$ bound.
Note that, to someone who only looks at $(\varepsilon_j, \delta)$ to describe the budget, \textbf{this attack goes completely unnoticed} as the privacy budget is observed (at least at the calibration point).

Thus, while the privacy guarantee for $x_j$ is unchanged, its specific sampling rate and noise multiplier are determined to increase the sample-level vulnerability.
Furthermore, the attacker can tune the choice in per-sample privacy budgets to limit the corresponding performance degradation, consequently trading off the advantage with the utility cost (and thus fingerprint) of this attack.
This strategy has theoretical and empirical support, as evidenced in the previous sections and the empirical evaluation in \autoref{sec:eval_attack}.


Consider a standard machine learning training scenario in which data is collected from individual contributors. 
Each contributor establishes a contract with the data-collecting entity, specifying that their data may only be used in a manner that expands a personalized privacy budget, denoted as $\varepsilon_j^{max}, \delta$. 
A model trainer, tasked with training a machine learning model, must ensure that this contract is respected.

Notably, the contract defines only an upper bound on the privacy loss for each contributor. 
This leaves the model trainer free to select any privacy budget for each user that does not exceed the contractually specified upper bound. Consequently, if the trainer knows these bounds, they can exploit this flexibility to choose privacy budgets strategically. 
Specifically, using the sampling-based iDP mechanism, the model could be trained to maximize the privacy risk for a targeted individual while still technically remaining within the contract limits. 
This illustrates that the proposed threat model is especially realistic: even when formal contracts or guarantees exist, sampling-based iDP opens a pathway for adversaries to selectively increase the exposure risk of specific individuals.
Moreover, it solely requires a centralized instance coordinating the iDP training, common in machine learning settings.

Remark: While the attack surfaces excess vulnerabilities of the target data point, it does not extract information about it.
Nonetheless, it makes extracting information easier for any subsequent attack, without imposing any constraints on this subsequent attack.

\subsection{Collusion Attack}

\begin{figure}[htbp]
    \centering
    \includesvg[width=0.25\textwidth]{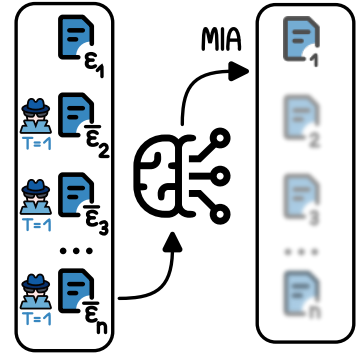} 
    \caption{Concept visualization of the collusion attack. Individual data contributors collude on a chosen target (T=1) and choose privacy budgets ($\bar{\varepsilon}_2, \bar{\varepsilon}_3, \dots \bar{\varepsilon}_n$) to increase the vulnerability of the target data.}
    \label{fig:attack2}
\end{figure}

Building on the concept of the \textit{Budget Manipulation Attack}, we now consider a more potent threat. 
In many practical systems, central administrators are considered trusted entities that must obey the privacy budgets chosen by each data contributing individual based on their privacy preferences. 
While this empowers data contributors, it also opens the door to a new vulnerability: malicious collusion. 

This attack explores a scenario where a group of data holders collude against another user. 
They can collectively surface a targeted individual's vulnerabilities by coordinating and strategically manipulating their privacy demands. 
This attack is particularly realistic as it requires no central authority, emerging organically from the actions of the users themselves. 
The attack is visualized in \autoref{fig:attack2} and the threat model is formally defined as follows:

\begin{definition}[Threat Model: Adversarial Collusion]
Let $\mathcal{A}$ be a training algorithm that satisfies iDP. 
For a dataset $D = \{x_1, \dots, x_n\}$ contributed by $n$ distinct data holders, the algorithm $\mathcal{A}$ provides $(\varepsilon_i, \delta)$-DP for each sample $x_i$ with respect to its corresponding privacy budget $\varepsilon_i$.

The adversary is now a coalition of data holders manipulating their privacy budgets to target a specific individual. 
Their capabilities and objectives are defined as follows:
\begin{enumerate}
    \item \textbf{No Model Access.} The adversary has no direct access to the trained model $\mathcal{A}(D)$, its parameters, or any intermediate computations. The attack is conducted entirely before the training phase by manipulating privacy budgets.
    
    \item \textbf{Adversarial Collusion and Target Selection.} A subset of data holders forms an adversarial coalition, denoted by the index set $C \subset \{1, \dots, n\}$. This coalition coordinates to select a target individual, indexed by $j \in \{1, \dots, n\} \setminus C$, to compromise their privacy.
    
    \item \textbf{Decentralized and Self-Controlled Privacy Budgets.} Each member $i \in C$ of the adversarial coalition has complete control over their own privacy budget, $\varepsilon_i$. 
    The coalition members cannot directly alter the budgets of the target, $\varepsilon_j$, or any non-colluding data holders, $\{\varepsilon_k\}_{k \notin C \cup \{i\}}$.
    
    \item \textbf{Coordinated Attack Objective.} The members of the coalition $C$ coordinate to choose their respective privacy budgets, $\{\varepsilon_i\}_{i \in C}$, in a way that maximizes the privacy vulnerability of the target individual $j$. Assuming the budgets of non-adversarial parties are fixed, the coalition solves the following optimization problem:
    \[
    \max_{\{\varepsilon_i\}_{i \in C}} \mathbb{E}_{\mathcal{A}} \left[ V(x_j, \mathcal{A}(D)) \right],
    \]
    where $V(x_j, \mathcal{A}(D))$ is a function that quantifies the privacy risk or vulnerability of the target's data point $x_j$ (e.g., the success probability of a membership inference attack against $x_j$). The expectation is taken over the randomness inherent in the training algorithm $\mathcal{A}$.
\end{enumerate}
\end{definition}

In this context, data holders may correspond either to individual contributors or to entire data silos, which makes the threat model particularly relevant in federated learning scenarios, as multiple data silos can collude to increase the excess privacy risk of the data in a targeted silo. 
For instance, if data is equally distributed across five silos, and four of them collude, then 80\% of the total privacy budgets can be strategically manipulated to increase the exposure risk of the fifth silo's data. 
Crucially, this occurs without any influence or control from the data holders of the targeted silo or a central administrator, emphasizing how collusion in sampling-based iDP federated settings can severely undermine individual or institutional privacy guarantees.

As with the budget manipulation attack, a promising attack strategy in this threat model is for the colluding group of adversaries to choose minimal individual privacy budgets.

Both theoretical considerations and empirical findings, presented in the previous sections and \autoref{sec:eval_attack}, substantiate this strategy. Most importantly, previous sections have shown that as much as 20\% of colluding data points significantly affect the excess privacy risk of the remaining data.

\subsection{Evaluating Attacks}
\label{sec:eval_attack}
Although the two attacks introduced earlier differ in their formal definitions, they share a common implication for model training: both aim to increase the vulnerability of a specific target sample by manipulating the allocation of privacy budgets.
In both scenarios, the adversary follows a similar strategy, whether a centralized authority or a colluding group of data owners. 
They assign \textit{minimal} privacy budgets to all non-targeted data points, while allocating the \textit{maximal allowable} privacy budget to the target sample. 
This influences the parameters of the privacy-preserving mechanism to increase the excess vulnerability of the target data point.
While a colluding group of users can only influence the budgets of the samples they contribute, a central adversary can arbitrarily set the privacy budgets for all data points in the training set. 
To avoid the detection of malicious behaviour, the attacks may be limited to operating within specific bounds of utility degradation induced by lower than allowed privacy budgets for a substantial portion of contributed data.

\begin{figure}
    \centering
    \includesvg[width=0.9\linewidth]{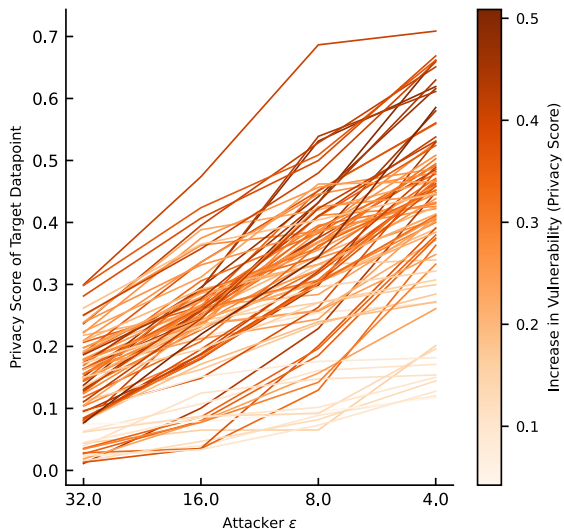}
    \caption{This spaghetti plot illustrates the increase in sample-level MIA susceptibility as described by the privacy score in an adversarial setting. 
    Each line represents a single attacked data point. 
    The plot shows the results for a group of attackers who collude to decrease their privacy budget from $\epsilon=32$ to $\epsilon=4$ at $\delta=10^{-12}$ on the Credit Card Default dataset. This reduction in privacy budget surfaces excess vulnerabilities on the data of the victim, which is empirically shown by the higher privacy scores on their data (y-axis). 
    A colour map is used to provide better visibility of how much samples increase in vulnerability from attacker privacy budget $\epsilon=4$ to attacker privacy budget $\epsilon=32$. 
    Due to the \textit{privacy onion effect}, only some samples suffer higher vulnerability, where adversarial attacks are largely successful ($62\%$ of the data). 
    Samples where the attack has no effect (neither an increase nor a decrease in privacy scores) are not shown for visualization.}
    \label{fig:budget_man_attack}
\end{figure}

To evaluate the attacks empirically, we need to perform a subsequent attack as LiRA-based MIA for $m$ investigated data points.
We perform MIA across $b$ investigated privacy budget distributions, each requiring training $n$ distinct shadow models, for each target sample.
As we are only interested in the change in vulnerability of a \textit{single} targeted data point, the IN and OUT datasets for LiRA differ solely in including this single target data point.
This makes LiRA more expressive as it avoids reasoning about single data points using aggregates of the subsampled datasets.
Consequently, $n = 64$ yields sufficiently robust estimates of the MIA susceptibility of individual data points.
While training on the full dataset, we limit the evaluation to $1000$ random samples.
However, this still requires us to train $m\times b\times n = 1000 \times 4 \times 64 = 256000$ shadow models. 
Consequently, we limit the evaluation to the more powerful \textit{budget manipulation attack} in \autoref{fig:budget_man_attack} exemplified on the credit card default dataset.
We expect the result of the collusion attack to scale with the proportion of the size of the colluding party relative to the dataset, as indicated in previously established theoretical analysis (\autoref{fig:privacy_tradeoff}, \autoref{fig:3d_privacy_tradeoff}), and experiments (\autoref{tab:empirical_priv_images}).
\autoref{fig:budget_man_attack} shows the privacy scores ($\mathrm{priv}$) on targeted data points in the \textit{budget manipulation attack} when the adversary manipulates all privacy budgets except for the target data point.
Our example assumes a uniform privacy budget of $([32, 16, 8, 4], 10^ {- 12})$ across all non-target data points and $(32, 10^{-12})$ for the target data point.

When a machine learning model is trained with adversarially selected privacy budgets in sampling-based iDP, the vulnerability of the targeted data point increases.
Since both Budget Manipulation and Collusion attacks operate in a similar way, the following evaluation applies to both.
\autoref{fig:budget_man_attack} shows that for a substantial subset of the investigated samples, this attack leads to a notable increase in vulnerability with an increase in privacy score of up to $0.5$.
While the attack is highly effective for this subset, some of the investigated samples are not influenced by the attack. 
This is justified by the \textit{privacy onion effect} \citep{carlini2022privacy}, with many samples having reduced individual susceptibility to extraction.
Nonetheless, we argue that the attack remains highly concerning because it operates entirely within the formal guarantees of DP and dramatically increases the risk of a large proportion of individuals. 
The increased vulnerability is, in principle, accepted by the data contributor when consenting to a given privacy budget. 
However, this risk remains unreported within the \epsdel-DP framework, as \epsdel-DP does not explicitly capture this increase in vulnerability. 

\section{Bounding Excess Vulnerability}
The emergence of these excess vulnerabilities is fundamentally based on the incomplete description of the privacy protection in terms of an \epsdel tuple.
As shown both in previous works \citep{kaissis2024beyond, hayes2023bounding} and our analysis, providing an \epsdel-iDP guarantee to a data contributor does not provide holistic information about their privacy risk.

\begin{figure}[htbp]
    \centering
    \includesvg[width=0.8\linewidth]{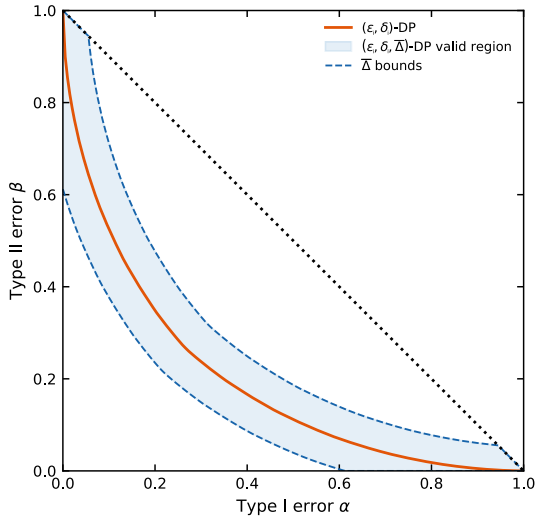} 
    \caption{The blue area denotes the valid region for an $(8, 10^{-5}, 0.05)$-DP guarantee, bounded by mechanisms that are $\Delta$-approximately Blackwell-dominant over a baseline $(8, 10^{-5})$-DP mechanism (a subsampled Gaussian mechanism with a sample rate of $0.1$, noise multiplier of $0.66$, and $10$ iterations).}
    \label{fig:Delta_bound}
\end{figure}

A trivial bound on the excess vulnerability could be established by bounding the maximal adversarial advantage. However, the computation of the advantage is based on the assumption of a fair coin flip probability of the inclusion of a sample, and thus changes substantially depending on the prior information of the adversary \citep{kaissis2024beyond}.
Therefore, we propose to expand the \epsdel formulation of differential privacy to \epsdeldel to bound the privacy risk of all candidate mechanisms directly with a pair of approximately Blackwell dominant mechanisms, with $\Delta$ defined as \citep{kaissis2024beyond}:

\begin{definition}[$\Delta$-Divergence \citep{kaissis2024beyond}]
The $\Delta$-divergence of a mechanism $\mathcal{M}$ to another mechanism $\widetilde{\mathcal{M}}$ is given by:
\[
\Delta(\mathcal{M} \,\|\, \widetilde{\mathcal{M}}) = \inf \{ \kappa \ge 0 \mid \forall \alpha : f(\alpha + \kappa) - \kappa \le \tilde{f}(\alpha) \}
\]
where $f$ and $\tilde{f}$ are the trade-off functions of $\mathcal{M}$ and $\widetilde{\mathcal{M}}$, respectively. In essence, the $\Delta$-divergence provides a notion of approximate Blackwell dominance (privacy guarantee ranking up to a small constant representing the maximum admissible excess vulnerability over all adversarial priors), allowing for a robust comparison of mechanisms based on their complete privacy profiles.
\end{definition}
Intuitively, the $\Delta$-divergence measures a mechanism's increase in excess vulnerability over a baseline mechanism, providing a hard upper bound on excess MIA vulnerability.
To bound this excess vulnerability, we propose to construct the \epsdeldel-DP formulation with $\overline{\Delta}$ providing an upper and lower bound on the trade-off curve centered around the trade-off curve of a standard DP mechanism, assuming a single privacy budget group with budget \epsdel.
Consequently, a valid mechanism under an \epsdeldel guarantee is defined as:

\begin{definition}[\epsdeldel-iDP]
Let $\mathcal{M}$ be a standard DP mechanism where all samples share a single privacy budget \epsdel. 
Let $\widetilde{\mathcal{M}}$ be a mechanism that differs from $\mathcal{M}$ by using sample-level individualized privacy budgets (iDP).
We say that $\widetilde{\mathcal{M}}$ satisfies \textbf{\epsdeldel-iDP} for $x_i$ if its symmetric $\Delta$-divergence from $\mathcal{M}$ with $\varepsilon=\varepsilon_i$, $\delta=\delta_i$ is bounded by~$\overline\Delta$:
\[
\Delta^{\leftrightarrow} = \max\left(\Delta(\mathcal{M} \,\|\, \widetilde{\mathcal{M}}),  \Delta(\widetilde{\mathcal{M}}\,\|\,\mathcal{M})\right) \leq \overline\Delta,
\]
with $\Delta^{\leftrightarrow}$ as defined in \citet{kaissis2024beyond}.
\end{definition}
$\Delta^{\leftrightarrow}$ of a set of mechanisms can be computed in negligible time \cite{kaissis2024beyond}.
The \epsdeldel-DP guarantee ensures that the privacy risk of an iDP mechanism does not deviate substantially from the risk profile of the standard, non-individualized mechanism that a data contributor would typically expect. 
Intuitively, $\overline\Delta$ acts as a \textit{stability guarantee}, assuring any user that their actual privacy risk will not deviate from a standard, non-individualized baseline by more than a pre-specified amount, regardless of the choices made by other users.
This approach effectively bounds the excess risk while still providing enough flexibility (\textit{slack}) for mechanism design, to enable practical, sampling-based iDP methods. 
A visualization of an \epsdeldel allowable region is provided in \autoref{fig:Delta_bound}.
While \epsdeldel-iDP allows data contributors to establish a contract that bounds their excess risk, it can also inform mechanism design.
A potential meta-algorithm could be described as:
\textit{(1)} If the slack in mechanism design allowed for in the \epsdeldel-iDP formulation allows for sampling-based iDP, then employ sampling-based iDP for higher utility than other methods.
\textit{(2)} If sampling-based iDP would introduce excess vulnerability outside of the $\overline\Delta$ bound, either pad the dataset with public data (see \citet{nasr2023effectively}), assigning it a specific privacy budget to control the privacy budget distribution, or employ sensitivity-based iDP. 
Thus, \epsdeldel prevents mechanism parameter choices that result in a privacy profile diverging too far from a baseline. 
Rejecting a mechanism because of a high $\Delta^{\leftrightarrow}$, the model achieved this \say{better utility} by allowing impermissibly high information leakage. 
\epsdeldel enforces an \say{honest} privacy-utility trade-off, preventing utility gains that are effectively subsidized by compromised privacy.

\section{Discussion and Conclusion}
Our work reveals a subtle yet critical vulnerability in sampling-based iDP: the promise of individual control is compromised by the systemic interdependence of all data contributors' privacy preferences. 
In sampling-based iDP \citep{boenisch2023have}, an individual's privacy risk is not solely determined by their chosen ($\varepsilon$, $\delta$)-budget, but also by the dataset-wide distribution of privacy budgets. 
This gives rise to \say{excess vulnerability} where mechanisms formally satisfy every user's iDP guarantee, but still expose users to greater unreported risks.

Users who choose a weak privacy budget (a high $\varepsilon$) may paradoxically become even more vulnerable if they are in the minority, among users with stronger privacy preferences. 
This fundamentally challenges the premise of sampling-based iDP as a purely personal privacy control, revealing it instead as a system with unstated collective dynamics.
Furthermore, this highlights the importance of carefully considering implementation details in individualized privacy control.

Building on this core vulnerability, we introduced two attacks: the Budget Manipulation Attack and the Adversarial Collusion Attack. 
While the first aligns well with standard machine learning settings with a central coordinator, the latter is particularly concerning as it presents a decentralized threat. 
and can emerge organically from a group of users coordinating their actions. 
By strategically selecting privacy budgets, a central administrator, or a colluding group of adversaries, can amplify a targeted individual's vulnerability without violating the ($\varepsilon_i$,$\delta_i$)-based promise made to the targeted individual. 
These attacks are subtle and feasible, operating entirely within the formal iDP bounds, unlike traditional attacks that seek to \textit{break} the DP-guarantee or require privileged access.
Our findings necessitate a reevaluation of how we design and audit iDP systems, as we demonstrate that it is insufficient to verify that a mechanism satisfies a set of individual ($\varepsilon_i$, $\delta_i$) constraints. 
Our empirical validation used LiRA for MIA, with MIA success mathematically implying tight bounds on attribute inference, singling out, and data reconstruction \citep{kulynych2025unifying}.

While the excess vulnerability highlighted by our work is inherent to sampling-based iDP, MIA is known to have limited capabilities in auditing privacy \citep{mia_critique_chowdhury2025reliability}, especially when scaled to LLMs \citep{mia_critique_2_zhang2025position}.
Consequently, further research is needed to understand how these vulnerabilities manifest in large-scale settings with diverse data modalities. 
The practical execution of a collusion attack also depends on the attackers' ability to coordinate and potentially accept a utility trade-off, which is strongly context and domain-dependent. 
Moreover, due to the privacy onion effect \citep{carlini2022privacy}, not all samples in a dataset expend all privacy budget during training.
Finally, while we propose a direction for mitigation by bounding excess vulnerability, developing a comprehensive and practical defence mechanism remains an open challenge.
While we propose to bond the excess risk using \epsdeldel-iDP, another possible option would be to employ the $f-DP$ formulation of DP, directly bounding the domain of trade-off functions and consequently excess risk. 
However, to allow for mechanism design, this also requires establishing a contract with the data contributor that allows for a range of trade-off functions and consequently requires a loose bound on the MIA advantage, while losing the (already limited \citep{cummings2021need}) interpretability of \epsdel-DP.

Future research should focus on the development of robust mitigation strategies and explore the socio-technical aspects of this vulnerability. 
How should users be informed of this interdependence? 
What new policies or user interface designs could help manage these collective risks? 
What is a formal definition of differential privacy that is interpretable for data contributors with respect to their personal risk?

In conclusion, this paper demonstrates that the move towards personalized privacy  through sampling-based iDP introduces new gameable vulnerabilities that necessitates rethinking of how privacy protection is communicated to data contributors to enable them to keep control over their privacy risks. 

\section*{Acknowledgements}
JK received support from the European Union under Grant Agreement 101100633 (EUCAIM). Views and opinions expressed are, however, those of the author(s) only and not necessarily reflect those of the European Union or the European Commission. Neither the European Union nor the granting authority can be held responsible for them.

AZ and DR received support from the Federal Ministry of Research, Technology and Space of Germany and the Medical Informatics Initiative as part of the PrivateAIM Project (grant no. 01ZZ2316C)

\section*{LLM usage considerations}
LLMs were used for editorial purposes in this manuscript, and all outputs were inspected by the authors to ensure accuracy and originality.

\bibliographystyle{plainnat}
\bibliography{bibliography}

\begin{thebibliography}{61}
\providecommand{\natexlab}[1]{#1}
\providecommand{\url}[1]{\texttt{#1}}
\expandafter\ifx\csname urlstyle\endcsname\relax
  \providecommand{\doi}[1]{doi: #1}\else
  \providecommand{\doi}{doi: \begingroup \urlstyle{rm}\Url}\fi

\bibitem[Abadi et~al.(2016)Abadi, Chu, Goodfellow, McMahan, Mironov, Talwar, and Zhang]{abadi2016deep}
Martin Abadi, Andy Chu, Ian Goodfellow, H~Brendan McMahan, Ilya Mironov, Kunal Talwar, and Li~Zhang.
\newblock Deep learning with differential privacy.
\newblock In \emph{Proceedings of the 2016 ACM SIGSAC conference on computer and communications security}, pages 308--318, 2016.

\bibitem[Acevedo et~al.(2020)Acevedo, Merino, Alférez, Ángel Molina, Boldú, and Rodellar]{ACEVEDO2020105474}
Andrea Acevedo, Anna Merino, Santiago Alférez, Ángel Molina, Laura Boldú, and José Rodellar.
\newblock A dataset of microscopic peripheral blood cell images for development of automatic recognition systems.
\newblock \emph{Data in Brief}, 30:\penalty0 105474, 2020.
\newblock ISSN 2352-3409.

\bibitem[Balle et~al.(2020)Balle, Barthe, and Gaboardi]{balle2020privacy}
Borja Balle, Gilles Barthe, and Marco Gaboardi.
\newblock Privacy profiles and amplification by subsampling.
\newblock \emph{Journal of Privacy and Confidentiality}, 10\penalty0 (1), 2020.

\bibitem[Berendt et~al.(2005)Berendt, Günther, and Spiekermann]{Berendt2005Privacy}
Bettina Berendt, Oliver Günther, and Sarah Spiekermann.
\newblock Privacy in e-commerce: Stated preferences vs. actual behavior.
\newblock \emph{Commun. ACM}, 48:\penalty0 101--106, 01 2005.

\bibitem[Boenisch et~al.(2023)Boenisch, M\"{u}hl, Dziedzic, Rinberg, and Papernot]{boenisch2023have}
Franziska Boenisch, Christopher M\"{u}hl, Adam Dziedzic, Roy Rinberg, and Nicolas Papernot.
\newblock Have it your way: individualized privacy assignment for dp-sgd.
\newblock In \emph{Proceedings of the 37th International Conference on Neural Information Processing Systems}, NIPS '23, Red Hook, NY, USA, 2023. Curran Associates Inc.

\bibitem[Boglioni et~al.(2025)Boglioni, Liu, Ilyas, and Wu]{boglioni2025optimizing}
Matteo Boglioni, Terrance Liu, Andrew Ilyas, and Zhiwei~Steven Wu.
\newblock Optimizing canaries for privacy auditing with metagradient descent.
\newblock \emph{arXiv preprint arXiv:2507.15836}, 2025.

\bibitem[Carlini et~al.(2022{\natexlab{a}})Carlini, Chien, Nasr, Song, Terzis, and Tramer]{carlini2022membership}
Nicholas Carlini, Steve Chien, Milad Nasr, Shuang Song, Andreas Terzis, and Florian Tramer.
\newblock Membership inference attacks from first principles.
\newblock In \emph{2022 IEEE symposium on security and privacy (SP)}, pages 1897--1914. IEEE, 2022{\natexlab{a}}.

\bibitem[Carlini et~al.(2022{\natexlab{b}})Carlini, Jagielski, Zhang, Papernot, Terzis, and Tramer]{carlini2022privacy}
Nicholas Carlini, Matthew Jagielski, Chiyuan Zhang, Nicolas Papernot, Andreas Terzis, and Florian Tramer.
\newblock The privacy onion effect: Memorization is relative.
\newblock \emph{Advances in Neural Information Processing Systems}, 35:\penalty0 13263--13276, 2022{\natexlab{b}}.

\bibitem[Chen et~al.(2025)Chen, Zhang, Guo, and Gao]{chen2025role}
Jing Chen, Manling Zhang, Mengyao Guo, and Ze~Gao.
\newblock The role of risk attitudes in shaping digital privacy preferences: evidence from a large-scale survey.
\newblock \emph{Humanities and Social Sciences Communications}, 12\penalty0 (1):\penalty0 1--10, 2025.

\bibitem[Chowdhury et~al.(2025)Chowdhury, Kong, and Chaudhuri]{mia_critique_chowdhury2025reliability}
Amrita~Roy Chowdhury, Zhifeng Kong, and Kamalika Chaudhuri.
\newblock On the reliability of membership inference attacks.
\newblock In \emph{2025 IEEE Conference on Secure and Trustworthy Machine Learning (SaTML)}, pages 534--549. IEEE, 2025.

\bibitem[Cole and Fanty(1991)]{isolet_54}
Ron Cole and Mark Fanty.
\newblock {ISOLET}.
\newblock UCI Machine Learning Repository, 1991.
\newblock {DOI}: https://doi.org/10.24432/C51G69.

\bibitem[Cummings et~al.(2021)Cummings, Kaptchuk, and Redmiles]{cummings2021need}
Rachel Cummings, Gabriel Kaptchuk, and Elissa~M Redmiles.
\newblock " i need a better description": An investigation into user expectations for differential privacy.
\newblock In \emph{Proceedings of the 2021 ACM SIGSAC Conference on Computer and Communications Security}, pages 3037--3052, 2021.

\bibitem[Deng(2012)]{deng2012mnist}
Li~Deng.
\newblock The mnist database of handwritten digit images for machine learning research.
\newblock \emph{IEEE Signal Processing Magazine}, 29\penalty0 (6):\penalty0 141--142, 2012.

\bibitem[Dong et~al.(2022)Dong, Roth, and Su]{dong2022gaussian}
Jinshuo Dong, Aaron Roth, and Weijie~J Su.
\newblock Gaussian differential privacy.
\newblock \emph{Journal of the Royal Statistical Society Series B: Statistical Methodology}, 84\penalty0 (1):\penalty0 3--37, 2022.

\bibitem[Dwork et~al.(2006)Dwork, McSherry, Nissim, and Smith]{dwork2006calibrating}
Cynthia Dwork, Frank McSherry, Kobbi Nissim, and Adam Smith.
\newblock Calibrating noise to sensitivity in private data analysis.
\newblock In \emph{Theory of Cryptography: Third Theory of Cryptography Conference, TCC 2006, New York, NY, USA, March 4-7, 2006. Proceedings 3}, pages 265--284. Springer, 2006.

\bibitem[Feldman and Zhang(2020)]{feldman2020neural}
Vitaly Feldman and Chiyuan Zhang.
\newblock What neural networks memorize and why: Discovering the long tail via influence estimation.
\newblock \emph{Advances in Neural Information Processing Systems}, 33:\penalty0 2881--2891, 2020.

\bibitem[Feldman and Zrnic(2021)]{feldman2022individual}
Vitaly Feldman and Tijana Zrnic.
\newblock Individual privacy accounting via a r\'{e}nyi filter.
\newblock In M.~Ranzato, A.~Beygelzimer, Y.~Dauphin, P.S. Liang, and J.~Wortman Vaughan, editors, \emph{Advances in Neural Information Processing Systems}, volume~34, pages 28080--28091. Curran Associates, Inc., 2021.

\bibitem[Hayes et~al.(2023)Hayes, Balle, and Mahloujifar]{hayes2023bounding}
Jamie Hayes, Borja Balle, and Saeed Mahloujifar.
\newblock Bounding training data reconstruction in dp-sgd.
\newblock \emph{Advances in neural information processing systems}, 36:\penalty0 78696--78722, 2023.

\bibitem[He et~al.(2015)He, Zhang, Ren, and Sun]{ResNet}
Kaiming He, Xiangyu Zhang, Shaoqing Ren, and Jian Sun.
\newblock Deep residual learning for image recognition, 2015.
\newblock URL \url{https://arxiv.org/abs/1512.03385}.

\bibitem[Heo et~al.(2023)Heo, Seo, and Whang]{heo2023personalized}
Geon Heo, Junseok Seo, and Steven~Euijong Whang.
\newblock Personalized dp-sgd using sampling mechanisms, 2023.
\newblock URL \url{https://arxiv.org/abs/2305.15165}.

\bibitem[Hisamoto et~al.(2020)Hisamoto, Post, and Duh]{hisamoto2020membership}
Sorami Hisamoto, Matt Post, and Kevin Duh.
\newblock Membership inference attacks on sequence-to-sequence models: Is my data in your machine translation system?
\newblock \emph{Transactions of the Association for Computational Linguistics}, 8:\penalty0 49--63, 2020.

\bibitem[Hofmann(1994)]{german_credit}
Hans Hofmann.
\newblock {Statlog (German Credit Data)}.
\newblock UCI Machine Learning Repository, 1994.
\newblock {DOI}: https://doi.org/10.24432/C5NC77.

\bibitem[Hui et~al.(2021)Hui, Yang, Yuan, Burlina, Gong, and Cao]{hui2021practical}
Bo~Hui, Yuchen Yang, Haolin Yuan, Philippe Burlina, Neil~Zhenqiang Gong, and Yinzhi Cao.
\newblock Practical blind membership inference attack via differential comparisons.
\newblock \emph{arXiv preprint arXiv:2101.01341}, 2021.

\bibitem[Jensen et~al.(2005)Jensen, Potts, and Jensen]{Jensen2005Privacy}
Carlos Jensen, Colin Potts, and Christian Jensen.
\newblock Privacy practices of internet users: Self-reports versus observed behavior.
\newblock \emph{International Journal of Human-Computer Studies}, 63\penalty0 (1):\penalty0 203--227, 2005.
\newblock ISSN 1071-5819.
\newblock HCI research in privacy and security.

\bibitem[Jorgensen et~al.(2015)Jorgensen, Yu, and Cormode]{jorgensen2015conservative}
Zach Jorgensen, Ting Yu, and Graham Cormode.
\newblock Conservative or liberal? personalized differential privacy.
\newblock In \emph{2015 IEEE 31St international conference on data engineering}, pages 1023--1034. IEEE, 2015.

\bibitem[Kaissis et~al.(2024)Kaissis, Kolek, Balle, Hayes, and Rueckert]{kaissis2024beyond}
Georgios Kaissis, Stefan Kolek, Borja Balle, Jamie Hayes, and Daniel Rueckert.
\newblock Beyond the calibration point: mechanism comparison in differential privacy.
\newblock In \emph{Proceedings of the 41st International Conference on Machine Learning}, ICML'24. JMLR.org, 2024.

\bibitem[Kan(2023)]{kan2023seeking}
Kazutoshi Kan.
\newblock Seeking the ideal privacy protection: strengths and limitations of differential privacy.
\newblock \emph{Monetary and Economic Studies}, 41:\penalty0 49--80, 2023.

\bibitem[Kasiviswanathan et~al.(2011)Kasiviswanathan, Lee, Nissim, Raskhodnikova, and Smith]{kasiviswanathan2011can}
Shiva~Prasad Kasiviswanathan, Homin~K Lee, Kobbi Nissim, Sofya Raskhodnikova, and Adam Smith.
\newblock What can we learn privately?
\newblock \emph{SIAM Journal on Computing}, 40\penalty0 (3):\penalty0 793--826, 2011.

\bibitem[Kermany et~al.(2018)Kermany, Goldbaum, Cai, Valentim, Liang, Baxter, McKeown, Yang, Wu, Yan, Dong, Prasadha, Pei, Ting, Zhu, Li, Hewett, Dong, Ziyar, Shi, Zhang, Zheng, Hou, Shi, Fu, Duan, Huu, Wen, Zhang, Zhang, Li, Wang, Singer, Sun, Xu, Tafreshi, Lewis, Xia, and Zhang]{KERMANY20181122}
Daniel~S. Kermany, Michael Goldbaum, Wenjia Cai, Carolina~C.S. Valentim, Huiying Liang, Sally~L. Baxter, Alex McKeown, Ge~Yang, Xiaokang Wu, Fangbing Yan, Justin Dong, Made~K. Prasadha, Jacqueline Pei, Magdalene~Y.L. Ting, Jie Zhu, Christina Li, Sierra Hewett, Jason Dong, Ian Ziyar, Alexander Shi, Runze Zhang, Lianghong Zheng, Rui Hou, William Shi, Xin Fu, Yaou Duan, Viet~A.N. Huu, Cindy Wen, Edward~D. Zhang, Charlotte~L. Zhang, Oulan Li, Xiaobo Wang, Michael~A. Singer, Xiaodong Sun, Jie Xu, Ali Tafreshi, M.~Anthony Lewis, Huimin Xia, and Kang Zhang.
\newblock Identifying medical diagnoses and treatable diseases by image-based deep learning.
\newblock \emph{Cell}, 172\penalty0 (5):\penalty0 1122--1131.e9, 2018.
\newblock ISSN 0092-8674.

\bibitem[Koskela et~al.(2023)Koskela, Tobaben, and Honkela]{koskela2022individual}
Antti Koskela, Marlon Tobaben, and Antti Honkela.
\newblock Individual privacy accounting with gaussian differential privacy, 2023.
\newblock URL \url{https://arxiv.org/abs/2209.15596}.

\bibitem[Krizhevsky(2009)]{krizhevsky2009learning}
Alex Krizhevsky.
\newblock Learning multiple layers of features from tiny images.
\newblock Technical Report TR-2009, University of Toronto, Toronto, ON, Canada, 2009.

\bibitem[Kulynych et~al.(2025)Kulynych, Gomez, Kaissis, Hayes, Balle, Calmon, and Raisaro]{kulynych2025unifying}
Bogdan Kulynych, Juan~Felipe Gomez, Georgios Kaissis, Jamie Hayes, Borja Balle, Flavio du~Pin Calmon, and Jean~Louis Raisaro.
\newblock Unifying re-identification, attribute inference, and data reconstruction risks in differential privacy.
\newblock \emph{arXiv preprint arXiv:2507.06969}, 2025.

\bibitem[Lokna et~al.(2023)Lokna, Paradis, Dimitrov, and Vechev]{lokna2023group}
Johan Lokna, Anouk Paradis, Dimitar~I Dimitrov, and Martin Vechev.
\newblock Group and attack: Auditing differential privacy.
\newblock In \emph{Proceedings of the 2023 ACM SIGSAC Conference on Computer and Communications Security}, pages 1905--1918, 2023.

\bibitem[Lécuyer(2021)]{lécuyer2021practical}
Mathias Lécuyer.
\newblock Practical privacy filters and odometers with r\'enyi differential privacy and applications to differentially private deep learning, 2021.
\newblock URL \url{https://arxiv.org/abs/2103.01379}.

\bibitem[Mahloujifar et~al.(2024)Mahloujifar, Melis, and Chaudhuri]{mahloujifar2024auditing}
Saeed Mahloujifar, Luca Melis, and Kamalika Chaudhuri.
\newblock Auditing $ f $-differential privacy in one run.
\newblock \emph{arXiv preprint arXiv:2410.22235}, 2024.

\bibitem[McKenna et~al.(2025)McKenna, Huang, Sinha, Balle, Charles, Choquette-Choo, Ghazi, Kaissis, Kumar, Liu, et~al.]{mckenna2025scaling}
Ryan McKenna, Yangsibo Huang, Amer Sinha, Borja Balle, Zachary Charles, Christopher~A Choquette-Choo, Badih Ghazi, George Kaissis, Ravi Kumar, Ruibo Liu, et~al.
\newblock Scaling laws for differentially private language models.
\newblock \emph{arXiv preprint arXiv:2501.18914}, 2025.

\bibitem[Mironov(2017)]{Mironov_2017}
Ilya Mironov.
\newblock Rényi differential privacy.
\newblock In \emph{2017 IEEE 30th Computer Security Foundations Symposium (CSF)}, page 263–275. IEEE, August 2017.

\bibitem[Nasr et~al.(2021)Nasr, Songi, Thakurta, Papernot, and Carlin]{nasr2021adversary}
Milad Nasr, Shuang Songi, Abhradeep Thakurta, Nicolas Papernot, and Nicholas Carlin.
\newblock Adversary instantiation: Lower bounds for differentially private machine learning.
\newblock In \emph{2021 IEEE Symposium on security and privacy (SP)}, pages 866--882. IEEE, 2021.

\bibitem[Nasr et~al.(2023)Nasr, Mahloujifar, Tang, Mittal, and Houmansadr]{nasr2023effectively}
Milad Nasr, Saeed Mahloujifar, Xinyu Tang, Prateek Mittal, and Amir Houmansadr.
\newblock Effectively using public data in privacy preserving machine learning.
\newblock In \emph{International Conference on Machine Learning}, pages 25718--25732. PMLR, 2023.

\bibitem[Rogers et~al.(2016)Rogers, Roth, Ullman, and Vadhan]{rogers2016privacy}
Ryan~M Rogers, Aaron Roth, Jonathan Ullman, and Salil Vadhan.
\newblock Privacy odometers and filters: Pay-as-you-go composition.
\newblock \emph{Advances in Neural Information Processing Systems}, 29, 2016.

\bibitem[Sablayrolles et~al.(2019)Sablayrolles, Douze, Schmid, Ollivier, and J{\'e}gou]{sablayrolles2019white}
Alexandre Sablayrolles, Matthijs Douze, Cordelia Schmid, Yann Ollivier, and Herv{\'e} J{\'e}gou.
\newblock White-box vs black-box: Bayes optimal strategies for membership inference.
\newblock In \emph{International Conference on Machine Learning}, pages 5558--5567. PMLR, 2019.

\bibitem[Salem et~al.(2018)Salem, Zhang, Humbert, Berrang, Fritz, and Backes]{salem2018ml}
Ahmed Salem, Yang Zhang, Mathias Humbert, Pascal Berrang, Mario Fritz, and Michael Backes.
\newblock Ml-leaks: Model and data independent membership inference attacks and defenses on machine learning models.
\newblock \emph{arXiv preprint arXiv:1806.01246}, 2018.

\bibitem[Shokri et~al.(2017)Shokri, Stronati, Song, and Shmatikov]{shokri2017membership}
Reza Shokri, Marco Stronati, Congzheng Song, and Vitaly Shmatikov.
\newblock Membership inference attacks against machine learning models.
\newblock In \emph{2017 IEEE symposium on security and privacy (SP)}, pages 3--18. IEEE, 2017.

\bibitem[Sinha et~al.(2025)Sinha, Mesnard, McKenna, Liu, Choquette-Choo, Huang, Yu, Kaissis, Charles, Liu, et~al.]{sinha2025vaultgemma}
Amer Sinha, Thomas Mesnard, Ryan McKenna, Daogao Liu, Christopher~A Choquette-Choo, Yangsibo Huang, Da~Yu, George Kaissis, Zachary Charles, Ruibo Liu, et~al.
\newblock Vaultgemma: A differentially private gemma model.
\newblock \emph{arXiv preprint arXiv:2510.15001}, 2025.

\bibitem[Song and Mittal(2021)]{song2021systematic}
Liwei Song and Prateek Mittal.
\newblock Systematic evaluation of privacy risks of machine learning models.
\newblock In \emph{30th USENIX security symposium (USENIX security 21)}, pages 2615--2632, 2021.

\bibitem[Song et~al.(2013)Song, Chaudhuri, and Sarwate]{Song2013StochasticGD}
Shuang Song, Kamalika Chaudhuri, and Anand~D. Sarwate.
\newblock Stochastic gradient descent with differentially private updates.
\newblock \emph{2013 IEEE Global Conference on Signal and Information Processing}, pages 245--248, 2013.

\bibitem[Steinke et~al.(2023)Steinke, Nasr, and Jagielski]{steinke2023privacy}
Thomas Steinke, Milad Nasr, and Matthew Jagielski.
\newblock Privacy auditing with one (1) training run.
\newblock \emph{Advances in Neural Information Processing Systems}, 36:\penalty0 49268--49280, 2023.

\bibitem[Taylor(2003)]{taylor2003most}
Humphrey Taylor.
\newblock Most people are "privacy pragmatists" who, while concerned about privacy, will sometimes trade it off for other benefits, 2003.
\newblock A Harris Poll study.

\bibitem[Truex et~al.(2019)Truex, Liu, Gursoy, Yu, and Wei]{truex2019demystifying}
Stacey Truex, Ling Liu, Mehmet~Emre Gursoy, Lei Yu, and Wenqi Wei.
\newblock Demystifying membership inference attacks in machine learning as a service.
\newblock \emph{IEEE transactions on services computing}, 14\penalty0 (6):\penalty0 2073--2089, 2019.

\bibitem[Tschandl et~al.(2018)Tschandl, Rosendahl, and Kittler]{Tschandl_2018}
Philipp Tschandl, Cliff Rosendahl, and Harald Kittler.
\newblock The ham10000 dataset, a large collection of multi-source dermatoscopic images of common pigmented skin lesions.
\newblock \emph{Scientific Data}, 5\penalty0 (1), August 2018.
\newblock ISSN 2052-4463.

\bibitem[Wang et~al.(2019)Wang, Balle, and Kasiviswanathan]{wang2019subsampled}
Yu-Xiang Wang, Borja Balle, and Shiva~Prasad Kasiviswanathan.
\newblock Subsampled r{\'e}nyi differential privacy and analytical moments accountant.
\newblock In \emph{The 22nd international conference on artificial intelligence and statistics}, pages 1226--1235. PMLR, 2019.

\bibitem[Watson et~al.(2021)Watson, Guo, Cormode, and Sablayrolles]{watson2021importance}
Lauren Watson, Chuan Guo, Graham Cormode, and Alex Sablayrolles.
\newblock On the importance of difficulty calibration in membership inference attacks, 2021.
\newblock URL \url{https://arxiv.org/abs/2111.08440}.

\bibitem[Whitehouse et~al.(2023)Whitehouse, Ramdas, Rogers, and Wu]{whitehouse2023fully}
Justin Whitehouse, Aaditya Ramdas, Ryan Rogers, and Steven Wu.
\newblock Fully-adaptive composition in differential privacy.
\newblock In \emph{International Conference on Machine Learning}, pages 36990--37007. PMLR, 2023.

\bibitem[Yang et~al.(2023)Yang, Shi, Wei, Liu, Zhao, Ke, Pfister, and Ni]{Yang_2023}
Jiancheng Yang, Rui Shi, Donglai Wei, Zequan Liu, Lin Zhao, Bilian Ke, Hanspeter Pfister, and Bingbing Ni.
\newblock Medmnist v2 - a large-scale lightweight benchmark for 2d and 3d biomedical image classification.
\newblock \emph{Scientific Data}, 10\penalty0 (1), January 2023.
\newblock ISSN 2052-4463.

\bibitem[Ye et~al.(2022)Ye, Maddi, Murakonda, Bindschaedler, and Shokri]{ye2022enhanced}
Jiayuan Ye, Aadyaa Maddi, Sasi~Kumar Murakonda, Vincent Bindschaedler, and Reza Shokri.
\newblock Enhanced membership inference attacks against machine learning models.
\newblock In \emph{Proceedings of the 2022 ACM SIGSAC conference on computer and communications security}, pages 3093--3106, 2022.

\bibitem[Yeh(2009)]{default_of_credit_card_clients_350}
I-Cheng Yeh.
\newblock {Default of Credit Card Clients}.
\newblock UCI Machine Learning Repository, 2009.
\newblock {DOI}: https://doi.org/10.24432/C55S3H.

\bibitem[Yeom et~al.(2018)Yeom, Giacomelli, Fredrikson, and Jha]{yeom2018privacy}
Samuel Yeom, Irene Giacomelli, Matt Fredrikson, and Somesh Jha.
\newblock Privacy risk in machine learning: Analyzing the connection to overfitting.
\newblock In \emph{2018 IEEE 31st computer security foundations symposium (CSF)}, pages 268--282. IEEE, 2018.

\bibitem[Yu et~al.(2024)Yu, Kamath, Kulkarni, Liu, Yin, and Zhang]{yu2022individual}
Da~Yu, Gautam Kamath, Janardhan Kulkarni, Tie-Yan Liu, Jian Yin, and Huishuai Zhang.
\newblock Individual privacy accounting for differentially private stochastic gradient descent, 2024.
\newblock URL \url{https://arxiv.org/abs/2206.02617}.

\bibitem[Zarifzadeh et~al.(2024)Zarifzadeh, Liu, and Shokri]{zarifzadeh2024low}
Sajjad Zarifzadeh, Philippe Liu, and Reza Shokri.
\newblock Low-cost high-power membership inference attacks, 2024, 2024.
\newblock URL \url{https://arxiv.org/abs/2312.03262}.

\bibitem[Zhang et~al.(2025)Zhang, Das, Kamath, and Tram{\`e}r]{mia_critique_2_zhang2025position}
Jie Zhang, Debeshee Das, Gautam Kamath, and Florian Tram{\`e}r.
\newblock Position: Membership inference attacks cannot prove that a model was trained on your data.
\newblock In \emph{2025 IEEE Conference on Secure and Trustworthy Machine Learning (SaTML)}, pages 333--345. IEEE, 2025.

\bibitem[Zhu et~al.(2022)Zhu, Dong, and Wang]{zhu2022optimalaccountingdifferentialprivacy}
Yuqing Zhu, Jinshuo Dong, and Yu-Xiang Wang.
\newblock Optimal accounting of differential privacy via characteristic function, 2022.
\newblock URL \url{https://arxiv.org/abs/2106.08567}.

\end{thebibliography}

\appendix

\subsection{On the Informed Choice of the Noise Multiplier in Sampling-Based iDP}
\label{app:fixed_batch_size}
Our analysis thus far has coupled privacy parameters by \textbf{constraining the parameter search to maintain a fixed expected batch size}. This linking paradigm establishes a strict interdependence among the sampling probabilities, which in turn depend on the \textbf{distribution} of privacy budgets across the dataset. Specifically, in \autoref{alg:algo1}, the fixed batch size constraint is enforced by lines 4-9, which couples the sampling probabilities to the privacy budget distribution.

\begin{algorithm}[htbp]
\caption{Finding Sample Parameters}
\label{alg:algo1}
\begin{algorithmic}[1]
\REQUIRE Per-group target privacy budgets $\{\varepsilon_1, \dots, \varepsilon_P\}$, target $\delta$, iterations $I$, number of total data points $N$ and per-privacy group data points $\{|G_1|, \dots, |G_P|\}$.

\STATE \textbf{init} $\sigma_{sample} \gets \text{getNoise}(\varepsilon_1, \delta, q, l)$
\STATE \textbf{init} $\{q_1, \dots, q_P\}$ where for $p \in [P]$
\STATE $q_p \gets \text{getSampleRate}(\varepsilon_p, \delta, \sigma_{sample}, I)$

\color{gray}\WHILE{$q \not\approx \frac{1}{N} \sum_{p=1}^P |G_p|q_p$}
    \STATE $\sigma_{sample} \gets s_i \sigma_{sample}$ \COMMENT{scaling factor: $s_i < 1$}
\color{black}
    \FOR{$p \in [P]$}
        \STATE $q_p \gets \text{getSampleRate}(\varepsilon_p, \delta, \sigma_{sample}, I)$
    \ENDFOR
\color{gray}\ENDWHILE

\RETURN $\sigma_{sample}, \{q_1, \dots, q_P\}$
\end{algorithmic}
\end{algorithm}

In essence, constraining the parameter search to a fixed batch is solely a specific way of aggregating the different privacy budgets to yield parameters that are most similar to training with a single privacy budget.
However, a interdependence among sampling parameters persists even when this constraint is relaxed. 
The algorithm still relies on a single, shared noise multiplier, $\sigma_{\text{sample}}$. In the original formulation, $\sigma_{\text{sample}}$ is derived from the minimum per-sample privacy budget, $\varepsilon_i$, which requires the largest noise multiplier (based on the expected overall sampling probability). 
This $\sigma_{\text{sample}}$ is then scaled down to compute noise multipliers and sampling rates for all other privacy budgets to satisfy the target batch size.

Consequently, the interdependence arises due to the informed choice of a single $\sigma_{\text{sample}}$ based on the desired privacy budgets used for all data points.
This reliance on a single $\sigma_{\text{sample}}$ introduces a critical vulnerability. Consider a scenario where the privacy budget of a single sample is altered, and this new budget becomes the minimum across the dataset. 
The shared $\sigma_{\text{sample}}$ must now be recomputed to accommodate this new, most stringent privacy requirement. Because all privacy groups use this single $\sigma_{\text{sample}}$, their corresponding noise multipliers and sampling rates must also be adjusted. This adaptation fundamentally changes their privacy profiles, leading to \textbf{excess privacy risks}.

\begin{figure}
    \centering
    \includesvg[width=\linewidth]{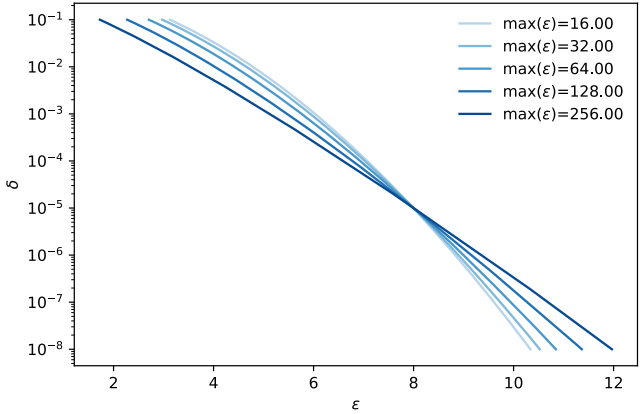}
    \caption{Privacy profiles for data points protected with $(8, 10^{-5})$-iDP when selecting $\sigma_{\text{sample}}$ based on the highest privacy budget in the dataset. While this has no interdependence with respect to a predefined expected batch size, each data point's sampling probability is computed to fit the global $\sigma_{\text{sample}}$.
    Consequently, when the value of the sample with the highest privacy budget changes, the sampling probability and thus privacy profiles need to change accordingly.
    With different privacy profiles, the samples are susceptible to different excess risk.}
    \label{fig:app_max_selection}
\end{figure}

Critically, this issue is not unique to a specific method of choosing $\sigma_{\text{sample}}$ (like by enforcing a predefined expected batch size) but is a systemic problem inherent to any \textbf{informed choice} of a shared noise multiplier. 
Different choices of $\sigma_{\text{sample}}$ lead to distinct types of privacy threats, as summarised in Table 1, which is a non-exhaustive list.
\autoref{fig:app_max_selection} compares the privacy profiles of samples with $(8, 10^{-5})$-iDP in the setting where $\sigma_{\text{sample}}$ is computed based on the sample with the largest $\varepsilon$.
Evidently, the privacy profiles change substantially, indicating varying excess risk.
Most importantly, this setting is especially critical as it requires just a single adversarial instance (getting assigned the highest privacy budget) to surface these excess risks for all data points in the dataset.

\begin{table}[t]
    \centering
    \caption{Summary of reasons for $\sigma_{\text{sample}}$ and their corresponding threats.}
    \label{tab:sigma_sample_threats}
    \begin{tabular}{ll}
        \toprule
        Reason for $\sigma_{\text{sample}}$ 
        & \raggedright\arraybackslash Threat \\
        \midrule
        $\mathrm{getNoise}\big(\min(\varepsilon_1, \dots, \varepsilon_p), \dots \big)$ 
        & A single sample controls \\ & the excess vulnerability \\
        $\mathrm{getNoise}\big(\max(\varepsilon_1, \dots, \varepsilon_p), \dots \big)$ 
        & A single sample controls \\ & the excess vulnerability \\
        $\mathrm{getNoise}\big(\mathrm{mean}(\varepsilon_1, \dots, \varepsilon_p), \dots \big)$ 
        & Interdependence; collective \\ & control of the excess vulnerability \\
        \bottomrule
    \end{tabular}
\end{table}

\subsection{Sensitivity-based iDP (Scale)}
\label{app:sens_based}

Sensitivity-based iDP (Scale) is an approach to achieving individualized privacy in DP-SGD. The core idea is to adjust the effective privacy guarantee for each data point by assigning it an individualized clipping norm rather than a uniform one. 
This method is computationally efficient because it does not require adding a different amount of noise per data point. 
Instead, a constant amount of noise is added, and the varying clip norms effectively change the signal-to-noise ratio for each user's contribution. Specifically, data points requiring a higher privacy budget (less privacy) are given a larger clip norm, which increases their gradient's magnitude relative to the added noise. 
Conversely, data points with a smaller budget (more privacy) receive a lower clip norm, which reduces their relative gradient magnitude, thereby providing a stronger privacy guarantee. 
This technique improves the overall utility of the final model by giving data points with higher privacy budgets a higher signal-to-noise ratio.
Finding the individual clipping bounds follows \autoref{alg:algo2}.

\begin{algorithm}[htbp]
\caption{Finding Scale Parameters}
\label{alg:algo2}
\begin{algorithmic}[1]
\REQUIRE Per-group target privacy budgets $\{\varepsilon_1, \dots, \varepsilon_P\}$, target $\delta$, number of iterations $I$, number of total data points $N$ and per-privacy group data points $\{|G_1|, \dots, |G_P|\}$, default clipping norm $c$, sample rate $q$
\STATE \textbf{init} $\{\sigma_1, \dots, \sigma_P\}$ where for $p \in [P]$:
\STATE $\sigma_p \leftarrow \text{getNoise}(\varepsilon_p, \delta, q, I)$
\STATE \textbf{init} $\sigma_{\text{scale}}: \sigma_{\text{scale}} \leftarrow (\frac{1}{N} \sum_{p=1}^P \frac{|G_p|}{c_p})^{-1}$ 
\FOR{$p \in [P]$}
    \STATE $c_p \leftarrow \frac{\sigma_{\text{scale}}c}{\sigma_p}$
\ENDFOR
\RETURN $\sigma_{scale}, \{c_1, \dots, c_P\}$
\end{algorithmic}
\end{algorithm}

\section{On Sensitivity-Based Individual Differential Privacy}
\label{app:sensitivit_based}
The sensitivity-based Individual Differential Privacy (iDP) framework achieves its privacy guarantees by assigning a unique, individualized clipping bound to each sample. This approach, as introduced by \citep{boenisch2023have}, utilizes a uniform noise multiplier, $\sigma_{\text{scale}}$, that is applied across all samples. This noise multiplier is elegantly derived as a weighted average of per-group noise multipliers, where the weights are proportional to the number of samples in each privacy group.

The global noise multiplier is formally defined as:
$$
\sigma_{\text{scale}} = \left( \frac{1}{N} \sum_{p=1}^P \frac{|\mathcal{G}_p|}{\sigma_p}\right)^{-1}
$$
where $N$ is the total number of samples in the dataset, $|\mathcal{G}_p|$ denotes the size of privacy group $p$, and $\sigma_p$ is the noise multiplier corresponding to the group's privacy budget $\varepsilon_p$.

Subsequently, the individualized clipping bound for each privacy group, $c_p$, is determined by the following relationship:
$$
c_p = \frac{\sigma_{\text{scale}}c}{\sigma_p}
$$
Here, $c$ represents the global gradient clipping norm.

In the canonical Differentially Private Stochastic Gradient Descent (DP-SGD) framework, the privacy expenditure and the resulting privacy profile are governed by the ratio of the applied noise to the algorithm's sensitivity. 
Under the sensitivity-based iDP paradigm, the privacy cost per step is dictated by the ratio $\sigma_{\text{scale}}/c_p$, which simplifies to:
$$
\frac{\sigma_{\text{scale}}}{c_p} = \frac{\sigma_{\text{scale}}}{\frac{\sigma_{\text{scale}}c}{\sigma_p}} = \frac{\sigma_p}{c}
$$
This fundamental result demonstrates that the theoretical privacy cost of sensitivity-based iDP precisely aligns with the privacy cost of the corresponding individual budgets under a non-individualized DP setting. 
Consequently, the signal-to-noise ratio for each sample during training remains theoretically identical to that in a standard DP-SGD configuration.

Despite this theoretical equivalence, the empirical behaviour may diverge. Using different clipping bounds and noise multipliers across samples can introduce variations in gradient bias, which may impact the algorithm's convergence and performance in practice.
Moreover, due to the varying privacy distribution, the learned representation changes as the model operates on different amounts of signal.

\begin{table*}[t!]
\centering
\caption{Test accuracy across different data splits for all datasets.}
\label{tab:accuracies}
\begin{tabular}{|l|c|c|c|c|c|c|}
\hline
\multirow{2}{*}{Data set} & \multirow{2}{*}{$\varepsilon_1$} & \multirow{2}{*}{$ \varepsilon_2$} &\multicolumn{4}{c|}{Test Accuracy} \\
\cline{4-7}
& & & 20\%/80\% & 40\%/60\% & 60\%/40\% & 80\%/20\% \\
\hline
Credit card default &  4 & 20 & $0.7198 \pm 0.0293$ & $0.7178 \pm 0.0305$ & $0.6914 \pm 0.0283$ & $0.6709 \pm 0.0183$ \\
German credit       &  4 & 16 & $0.7144 \pm 0.0302$ & $0.7084 \pm 0.0334$ & $0.6826 \pm 0.0261$ & $0.6674 \pm 0.0155$ \\
MNIST               &  4 & 16 & $0.8995 \pm 0.0033$ & $0.8959 \pm 0.0032$ & $0.8898 \pm 0.0030$ & $0.8843 \pm 0.0037$ \\
MNIST [4, 1000]     & 16 & 50 & $0.9706 \pm 0.0021$ & $0.9705 \pm 0.0021$ & $0.9696 \pm 0.0023$ & $0.9689 \pm 0.0023$ \\
UCI-Isolet          & 1  & 10 & $0.7351 \pm 0.0193$ & $0.7111 \pm 0.0185$ & $0.6580 \pm 0.0216$ & $0.5724 \pm 0.0228$ \\
CIFAR-10            & 16 & 50 & $0.4637 \pm 0.0069$ & $0.4577 \pm 0.0084$ & $0.4519 \pm 0.0080$ & $0.4427 \pm 0.0082$ \\
OrganCMNIST [2000]        &  8 & 32 & $0.7485 \pm 0.0067$ & $0.7391 \pm 0.0062$ & $0.7322 \pm 0.0076$ & $0.7233 \pm 0.0079$ \\
OrganSMNIST [2000]        &  8 & 32 & $0.5940 \pm 0.0081$ & $0.5858 \pm 0.0098$ & $0.5731 \pm 0.0031$ & $0.5631 \pm 0.0093$ \\
Pneumonia [2000]      &  8 & 32 & $0.8531 \pm 0.0130$ & $0.8559 \pm 0.0131$ & $0.8560 \pm 0.0156$ & $0.8508 \pm 0.0164$ \\
Ham10k [2000]      &  8 & 32 & $0.6883 \pm 0.0084$ & $0.6808 \pm 0.0096$ & $0.6832 \pm 0.01013$ & $0.6758 \pm 0.0128$ \\

\hline
\end{tabular}
\end{table*}

\subsection{\texorpdfstring{$\delta$}{δ} Calibration and Adversarial Advantage}
\label{app:delta}

As $\delta$ is shared across all privacy group proportions, it determines the calibration points, i.e., the point $(\varepsilon, \delta)$ to which we constrain our mechanism.
Calibrating to a different $\delta$ yields a different set of mechanisms, known to influence the adversarial advantage and consequently the privacy risk \cite{mckenna2025scaling}.
Our empirical analysis aims to investigate the effect of privacy budget distributions on the per-sample risks.
We deliberately set the calibration point to low $\delta$ values, i.e., $\delta=10^{-12}$.
The reasons, therefore, are twofold.

First, this describes an operationally interesting and realistic regime.
Recall that usually $\delta$ is set to be $\delta \ll 1/N$ with $N$ being the number of samples in the training dataset.
Consequently, with models, especially large language models (LLMs), being trained on increasingly large corpora of training data, this constitutes a realistic setting. 
Explicitly, VaultGemma \cite{sinha2025vaultgemma} a DP-LLM was trained with $\delta = 10^{-10}$.

\begin{figure}[htbp]
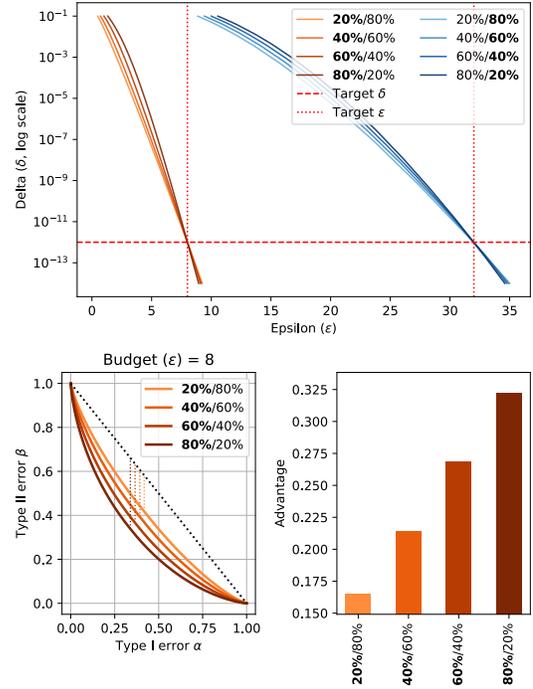

    \centering
    \includesvg[width=0.8\linewidth]{figures/appendix_delta/epsilon_delta_curve_2.svg} 
    \includesvg[width=0.8\linewidth]{figures/appendix_delta/trade_off_curves_2.svg} 
    \caption{Privacy profiles, trade-off functions, and adversarial advantage for a machine learning model trained with sampling-based iDP.
    While all mechanisms are calibrated to guarantee the predefined privacy budgets $\varepsilon_1 = 8, \varepsilon_2 = 32$ at $\delta=10^{-12}$ their advantage differs significantly  $0.165$ to $0.325$ for samples with $\varepsilon_1 = 8$}
    \label{fig:app_privacy_tradeoff_1}
\end{figure}

\begin{figure}[htbp]
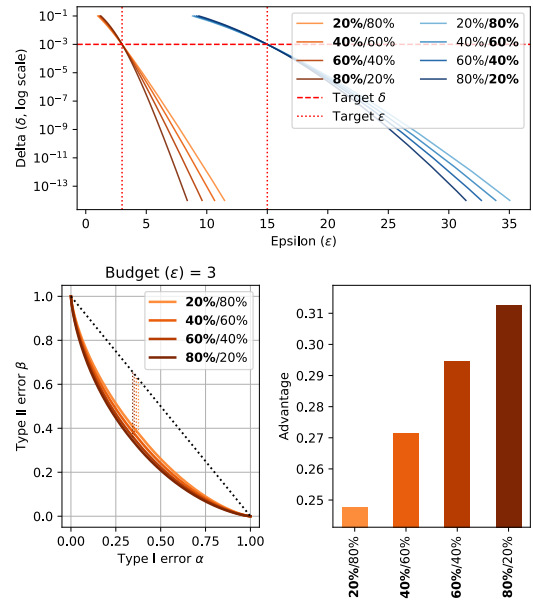

    \centering
    \includesvg[width=0.8\linewidth]{figures/appendix_delta/epsilon_delta_curve_1.svg} 
    \includesvg[width=0.8\linewidth]{figures/appendix_delta/trade_off_curves_1.svg} 
    \caption{Privacy profiles, trade-off functions, and adversarial advantage for a machine learning model trained with sampling-based iDP.
    While all mechanisms are calibrated to guarantee the predefined privacy budgets $\varepsilon_1 = 3, \varepsilon_2 = 15$ at $\delta=10^{3}$ their advantage differs significantly  $0.245$ to $0.315$ for samples with $\varepsilon_1 = 3$}
    \label{fig:app_privacy_tradeoff_2}
\end{figure}

Second, to empirically investigate the change in vulnerability when changing the privacy budget distributions, we deliberately chose a $\delta$ calibration point where the phenomenon we want to measure, i.e., the change in vulnerability, has a high magnitude. 
While it is difficult to make a general statement on how to calibrate $\delta$ for this particular target, our experiments evidence that decreasing $\delta$ increases the magnitude of the excess vulnerability.
Intuitively, vulnerability is measured at high $\delta$ values. 
By choosing a calibration (intersection) point of the privacy curves at a low $\delta$ we give the curves more \say{space} to diverge.
Comparing \autoref{fig:app_privacy_tradeoff_1} to \autoref{fig:app_privacy_tradeoff_2}, choosing a calibration point at $\delta=10^{-12}$ results in a change in adversarial advantage ranging from $0.165$ to $0.325$ ($\Delta_{diff}=0.16$) whereas calibrating the mechanisms to $\delta = 10^{-3}$ solely yields a adversarial advantage ranging from $0.245$ to $0.315$ ($\Delta_{diff}=0.07$) for different group compositions and $\varepsilon=3$ or $\varepsilon=8$ respectively.
In the compared scenarios, $(\varepsilon, \delta)$ is chosen to yield approximately equivalent privacy protection.

\subsection{Training Details}
\label{appendix:TrainingSettings}
\paragraph{Models}
We employ two types of neural architectures depending on the dataset modality: multilayer perceptrons (MLPs) for tabular and flattened image data, and simple convolutional neural networks (CNNs) or ResNet9 for image datasets with spatial structure. 
All models are deliberately kept small to reduce the computational cost of training the shadow models for LiRA.

The \textbf{MLP} architecture consists of a sequence of fully connected layers with ReLU activations. In its default form, the model takes the flattened input and projects it to a hidden layer of size 128, followed by another hidden layer of size 64, before mapping to the number of output classes.  

The \textbf{CNN} architecture is designed for small-scale image datasets. It uses two convolutional layers (with 32 and 64 kernels respectively, both of kernel size three and padding 1), each followed by ReLU activations and max-pooling operations. After flattening, the representation is passed through a fully connected layer of size 128, followed by the classification layer.  

The \textbf{ResNet-9} architecture follows the residual network design introduced by \citep{ResNet}. It is a lightweight variant of ResNet with only nine layers, making it well-suited for small-scale image datasets and faster experimentation while retaining the core idea of residual connections.

\paragraph{Datasets}
We evaluate our models on a diverse collection of datasets from computer vision, tabular classification benchmarks, and Kaggle competitions. Datasets are primarily selected from sensitive domains like finance (Credit Card Default; German Credit) or medicine (UCI Isolet, OrganCMNIST, OrganSMNIST, Peripheral Blood, Pneumonia, HAM10k). The main characteristics of the datasets are summarised in Table~\ref{tab:datasets}.

\begin{table}[ht]
\centering
\caption{Overview of datasets and associated model architectures. \textit{Input Dim.} refers to the feature dimensionality (flattened for images in case of MLPs, channel dimension for images in CNNs), and \textit{Spatial Dim.} is given for datasets with spatial structure.}
\label{tab:datasets}
\begin{adjustbox}{width=\linewidth}
\begin{tabular}{l l c c c}
\hline
\textbf{Dataset} & \textbf{Source} & \textbf{\#Classes} & \textbf{Input Dim.} & \textbf{Spatial Dim.} \\
\hline
Credit Card Default & \citep{default_of_credit_card_clients_350} & 2 & 59 & --  \\
German Credit & \citep{german_credit} & 2 & 59 & --  \\
MNIST & \citep{deng2012mnist} & 10 & 784 & -- \\
UCI Isolet & \citep{isolet_54} & 26 & 617 & --  \\
CIFAR-10 & \citep{krizhevsky2009learning} & 10 & 3 & 32  \\
OrganCMNIST & \citep{Yang_2023} & 11 & 1 & 28  \\
OrganSMNIST & \citep{Yang_2023} & 11 & 1 & 28 \\
Peripheral Blood & \citep{ACEVEDO2020105474} & 8 & 3 & 28  \\
Pneumonia & \citep{KERMANY20181122} & 2 & 1 & 28\\
HAM10k & \citep{Tschandl_2018} & 7 & 3 & 28  \\
\hline
\end{tabular}
\end{adjustbox}
\end{table}
Throughout the paper, a number following the dataset name indicates a reduced version of the dataset, with the number specifying the samples per class.
For example, the training set of OrganCMNIST [2000] consists of $11 \text{ (numbers of classes)} \times 2000 = 22000$ samples.
For MNIST [4, 1000] the $4$ indicates a reduction to the first four classes (i.e, 0, 1, 2, 3).
This reduces computational cost for training the shadow models.

\begin{table}[ht]
\centering
\caption{Overview of experimental settings.}
\label{tab:experiments}
\begin{adjustbox}{width=\linewidth}
\begin{tabular}{l l c c c}
\toprule
Dataset & Model & Learning Rate & Batch Size & Epochs \\
\midrule
Credit Card Default  & MLP      & 0.010 & 32  & 60 \\
German Credit        & MLP      & 0.010 & 32  & 60 \\
MNIST                & MLP      & 0.010 & 128 & 15 \\
UCI-Isolet           & MLP      & 0.010 & 256 & 25 \\
CIFAR-10             & CNN      & 0.010 & 128 & 15 \\
OrganCMNIST          & ResNet-9 & 0.005 & 128 & 25 \\
OrganSMNIST          & ResNet-9 & 0.005 & 128 & 25 \\
Peripheral Blood     & ResNet-9 & 0.005 & 128 & 25 \\
Pneumonia       & ResNet-9 & 0.005 & 128 & 25 \\
HAM10k               & ResNet-9 & 0.005 & 128 & 25 \\
\bottomrule
\end{tabular}
\end{adjustbox}
\end{table}

\paragraph{Experimental Details}
We employ LiRA based on $512$ shadow models ($128$ for the attack) to empirically evaluate the increase in MIA advantage.
All models (except for sensitivity-based iDP) were trained with differential privacy with a gradient clipping bound of $1$.
Privacy budgets are assigned to the samples randomly, such that the distribution of privacy budgets matches the desired privacy group sizes.
The budget-sample assignment remains consistent throughout all shadow models.
The shadow models for LiRA are trained on 50\% of the available dataset to generate IN and OUT logit distributions.
The split into IN and OUT data is facilitated, such that the privacy budget distribution remains as desired.
Moreover, across all 512 shadow models, each sample appears in approximately 50\% of the datasets.

Performing the split into IN and OUT datasets for all 512 models constrained such that a data point appears in approximately 50\% of model trainings (1) and that the privacy budget distribution remains unchanged (2) requires the computation of a \textbf{biregular binary assignment matrix}.
While the columns of the assignment matrix indicate the inclusion or exclusion of specific samples (columns) into the training of a specific shadow model (rows).
The biregular binary assignment matrix is constrained to have a fixed sum of column entries and a fixed sum of row entries.
By splitting the assignment matrix with respect to the privacy budgets and computing two separate biregular binary assignment matrices, we can enforce requirements (1) and (2).
As there is no non-trivial closed-form solution for constructing biregular binary assignment matrices, we approximate them while ensuring that the column sum exactly matches our desired privacy budget distribution and the row sum.
We weakly enforce a column sum of $n/2$ with $n$ being the number of shadow models.

All models were trained with the AdamW optimizer with weight decay of $5\times 10^{-4}$, Cross Entropy Loss and Cosine Annealing learning rate scheduling.
The training parameters are summarized in table \autoref{tab:experiments}.
The resulting accuracies are summarized in table \autoref{tab:accuracies}.


\end{document}